\documentclass[a4paper,12pt]{article}
\usepackage{jheppub} % for details on the use of the package, please
\usepackage{hyperref}
\hypersetup{
    bookmarks=true,         % show bookmarks bar?
    unicode=false,          % non-Latin characters in AcrobatÃÂÃÂÃÂÃÂ¢ÃÂÃÂÃÂÃÂÃÂÃÂÃÂÃÂs bookmarks
    pdftoolbar=true,        % show AcrobatÃÂÃÂÃÂÃÂ¢ÃÂÃÂÃÂÃÂÃÂÃÂÃÂÃÂs toolbar?
    pdfmenubar=true,        % show AcrobatÃÂÃÂÃÂÃÂ¢ÃÂÃÂÃÂÃÂÃÂÃÂÃÂÃÂs menu?
    pdffitwindow=false,     % window fit to page when opened
    pdfstartview={FitH},    % fits the width of the page to the window
    pdftitle={My title},    % title
    pdfauthor={Author},     % author
    pdfsubject={Subject},   % subject of the document
    pdfcreator={Creator},   % creator of the document
    pdfproducer={Producer}, % producer of the document
    pdfkeywords={keyword1} {key2} {key3}, % list of keywords
    pdfnewwindow=true,      % links in new window
    colorlinks=true,       % false: boxed links; true: colored links
    linkcolor=red,          % color of internal links
    citecolor=purple,        % color of links to bibliography
    filecolor=magenta,      % color of file links
    urlcolor=blue,           % color of external links
    linktocpage=true
}
\usepackage{graphics}
\usepackage{rotating}
\usepackage{subfigure}
\usepackage{pstricks}
\usepackage{color}
\usepackage{amsfonts}
\usepackage{mathrsfs,amsmath}
\usepackage{epsfig}
\usepackage{amsmath,amssymb,amsthm,graphicx,latexsym}
\usepackage{ulem}

\newcommand{\be}{\begin{equation}}
\newcommand{\ee}{\end{equation}}
\newcommand{\bea}{\begin{eqnarray}}
\newcommand{\eea}{\end{eqnarray}}
\newcommand{\bml}{\begin{subequations}}
\newcommand{\eml}{\end{subequations}}
\newcommand{\bfig}{\begin{figure}}
\newcommand{\efig}{\end{figure}}

\newcommand{\slashed}{\hspace{-1.1ex}/}

\newcommand{\bmat}{\begin{pmatrix}}
\newcommand{\emat}{\end{pmatrix}}
\begin{document}
	$~~~~~~~~~~~~~~~~~~~~~~~~~~~~~~~~~~~~~~~~~~~~~~~~~~~~~~~~~~~~~~~~~~~~~~~~~~~~~~~~~~~~$\textcolor{red}{\bf TIFR/TH/15-46}
	\title{\textsc{\fontsize{27}{90}\selectfont \sffamily \bfseries  Can Dark Matter be an artifact of extended theories of gravity?}}

	\author[a]{Sayantan Choudhury
		\footnote{\textcolor{purple}{\bf Presently working as a Visiting (Post-Doctoral) fellow at DTP, TIFR, Mumbai, \\$~~~~~$Alternative
				E-mail: sayanphysicsisi@gmail.com}. ${}^{}$}}
	\author[a]{Manibrata Sen}
	\author[b]{Soumya Sadhukhan}
	
	\affiliation[a]{Department of Theoretical Physics, Tata Institute of Fundamental Research, Colaba, Mumbai - 400005, India.
	}
	\affiliation[b]{Institute of Mathematical Sciences, C.I.T Campus, Taramani, Chennai - 600113, India.
	}
	\emailAdd{sayantan@theory.tifr.res.in, manibrata@theory.tifr.res.in , soumyasad@imsc.res.in}

	\abstract{In this article, we propose different background models of extended theories of gravity, which are minimally coupled to the SM fields, 
	to explain the possibility of genesis of dark matter without affecting the SM particle sector. We modify the gravity sector by allowing
	 quantum corrections motivated from (1) local $f(R)$ gravity and (2) non-minimally coupled gravity with SM sector and dilaton field. Next
	 we apply conformal transformation on the metric to transform the action back to the Einstein frame. We also show that an effective theory constructed from these extended theories of gravity and SM sector looks
	 exactly the same.
	 Using the relic constraint observed by Planck 2015, we constrain the scale of the effective field theory ($\Lambda_{UV}$) as well as the dark matter mass ($M$). We consider two cases- (1) light
	 dark matter (LDM) and (2) heavy dark matter (HDM), and deduce upper bounds on thermally
	 averaged cross section of dark matter annihilating to SM particles. Further we show that our model naturally incorporates self interactions of dark matter. Using these self interactions, we derive the constraints 
	 on the 
	 parameters of the (1) local $f(R)$ gravity and (2) non-minimally coupled gravity from dark matter self interaction.
	 Finally, we propose some different UV complete models from a particle physics point of view, which can give rise to the same effective theory that we have deduced from extended theories
	  of gravity.
	}

	%\begin{document} 
	\maketitle
	\flushbottom
	\section{Introduction}
	\label{aa1}
	
	Different cosmological measurements have confirmed that majority
	of the matter in this universe occurs in the form of a non-luminous 
	``dark matter''(DM). Infact DM accounts for almost $30\%$ 
	of the energy budget of the universe \cite{Komatsu}. Experimentally measured relic density of DM gives us some 
	insights into the particle nature of DM. It is a very well known fact
	that Standard Model (SM) of particle physics cannot provide any dark matter candidate. It is believed that to search for the existence of dark matter candidate, physics Beyond Standard Model (BSM)  is
	necessary \cite{Jungman,Bergstrom:2000pn,Feng:2010gw}. 
	These extensions of the SM are strongly motivated from observations of the galactic rotation curves, motion of galaxy clusters, two colliding clusters of galaxies in the 
	Bullet Cluster and cosmological observations \cite{Clowe:2006eq}. In such a scenario, the matter sector is modified without affecting the gravity sector. 
	But more precisely this type of approach is mostly ad-hoc as it does not always 
	provide any theoretical origin of such extensions in the matter sector (with the exceptions of a few DM models like neutralino WIMP, axion etc.).
	Alternatively these observations have also been explained through modification of the gravity sector without
	the need of any dark matter candidate, for 
	example:  Modified Newtonian dynamics (MOND) paradigm \cite{Milgrom:2009an} and Tensor-vector-scalar
	gravity (TeVeS) \cite{Bekenstein:2004ne}. But such proposals are not consistent with all the observational constraints \footnote{For example,
	MOND cannot completely eliminate the need for dark matter in astrophysical
	systems, since galaxy clusters show a residual mass discrepancy even when analyzed using MOND \cite{McGaugh:2014nsa} .}.
	To avoid the ambiguity of ad-hoc extensions of the SM, in this paper
	we propose an alternative framework based on the principles of
	Effective Field Theory (EFT) \cite{Busoni:2013lha,Busoni:2014sya,Busoni:2014haa,Macias:2015cna,Beltran:2008xg,Goodman:2010yf,Goodman:2010ku,Goodman:2010qn,Fan:2010gt,Cheung:2010ua,Cheung:2011nt,Cheung:2012gi,Fitzpatrick:2012ix,Kumar:2013iva,DeSimone:2013gj}. In this EFT approach, we represent the 
	interactions between DM and SM through a set of higher
	dimensional effective non-renormalizable Wilsonian operators, which are generated by integrating out the
	heavy mediator degrees of freedom at higher scales. This approach works best when there is a 
	clear separation of energy scales between the ultraviolet physics,
	and the relevant energy scales. This is
	clearly the case here, because when we consider indirect detection of DM, where two 
	DM particles annihilate to two SM particles, the momentum transferred
	in the process is of the order of the DM mass, which is clearly less
	than the energy scales considered. Even in case of direct detection, the momentum 
	transferred in a collision with a nuclei, is of the order of a few {\rm keV}. This justifies the use of an EFT.
	
	We start with the extended version of gravity sector keeping the SM matter sector unchanged. Such modifications in the gravity sector usually originates from quantum corrections in the gravity sector and are 
	motivated from various background higher dimensional field theoretic setups~\footnote{String Theory and its low energy versions provide such corrections in the gravity sector \cite{green1,green2,pol1,pol2}.
	Alternatively in ref.~\cite{love}, the author had shown that similar modifications in the gravity sector can be obtained form a geometrical perspective.}.
	One can also consider modification in the gravity sector by allowing non-minimal interaction between the matter field and  gravity~\footnote{In our case, the matter field is the scalar field which is 
	similar to the dilaton field appearing in scattering amplitudes of closed string theory \cite{green1,green2,pol1,pol2}. It is also important to note that, in the context of modified gravity, usually dilaton can be identified 
	to be the scalaron field \cite{Gorbunov:2010bn} originated from the higher curvature gravity sector.}. In the present context we use conformal transformation on the metric to explain the genesis of scalar
	dark matter from various types of extended 
	theories of gravity, i.e., local $f(R)$ gravity \cite{felice:2010,Sotiriou:2008rp}~\footnote{For eg., $f(R)$ gravity theory can explain the galaxy rotation curves \cite{Stabile:2013jon}.},
	  non-local theories of gravity \cite{Biswas:2011ar,Biswas:2012bp,Biswas:2013kla,Biswas:2013cha,Chialva:2014rla,Conroy:2014eja,Talaganis:2014ida,Conroy:2015wfa,Conroy:2015nva}~\footnote{In this work 
	  we have not discussed this possibility. We will report on this issue in our future work in this direction.} and finally
	we also allow non-minimal interaction between Einstein gravity with
	scalar matter field \cite{Bezrukov:2010jz,Bezrukov:2012hx,GarciaBellido:2012zu,GarciaBellido:2011de,Shaposhnikov:2009pv,Bezrukov:2009db,Bezrukov:2008ej,Bezrukov:2007ep,Choudhury:2013zna,Salvio:2015kka} as mentioned earlier.
	Thus in our prescribed methodology, although we start with an unchanged matter sector, it gets modified because of modifications in the gravity sector. This is where we differ from the contemporary ideas.
	Further to implement the constraint from observational probes~\footnote{Here we use Planck 2015 \cite{Ade:2015xua} data to constrain the relic density of dark matter.} on the relic density of the dark matter
	we use the tools and techniques of Effective Field Theory in the present setup. 
	
	Throughout the analysis of the paper we use the following sets of crucial assumptions:
	\begin{enumerate}
	 \item  We use the tools and techniques of the Effective Field Theory in the present context while applying the constraints from observational probes and indirect detection experiments.
	 Instead of introducing a Planckian cut-off at $M_{p}\sim 10^{19}~{\rm {\rm GeV}}$ here, we introduce a new UV cut-off scale, $\Lambda_{UV}<<M_{p}$ of the Effective Field Theory. In principle, more precisely this can be treated as 
	 the tuning parameter of the theoretical setup and we have shown explicitly from our prescribed analysis that this serves a very crucial role to satisfy the constraint for dark matter relic abundance
	 as obtained from Planck 2015 \cite{Ade:2015xua} data.
	 
	 \item We are implementing our prescribed methodology by taking some of the few well known examples of extended theories of gravity, i.e., local $f(R)$ gravity and non-minimally coupled gravity with 
	 scalar matter, in which, by applying conformal transformation on the metric one is able to construct a reduced and easier version of the theory in Einstein frame in terms of Einstein gravity, a new scalar 
	 matter field (dilaton) and an interaction between SM sector and dilaton 
	 matter field. In our prescription, we identify such a dilaton field to be the dark matter candidate.
	 
	 \item To validate to perturbative approximation appropriately in the present context we also assume that the interaction between SM sector and dilaton 
	 matter field is weak. Consequently, we expand the exponential dilaton coupling and due to large suppression by
	 the cut-off scale $\Lambda_{UV}$, we only take first three terms in the expansion series.
	 
	 \item Next we additionally impose a ${\cal Z}_{2}$ symmetry on dilaton, and drop the odd term under this symmetry.
	 As a result here we have only the first term ${\cal L}_{SM}$ and third term $\frac{\phi^2}{\Lambda^2_{UV}}{\cal L}_{SM}$. In our paper, the 
	 third term $\frac{\phi^2}{\Lambda^2_{UV}}{\cal L}_{SM}$ plays the significant role to describe the genesis of dilaton dark matter. One loop corrections to the dilaton mass puts an upper limit of
	 $M\leq 4\pi\Lambda_{UV}$ \cite{Blum:2014jca,Gannouji:2012iy}.

	 \item During our analysis we also assume that annihilation of DM at the galactic centre proceeds with a velocity  $v \approx 10^{-3}$.
             Consequently the thermally averaged cross-
section $\langle\sigma v\rangle$ is expanded in terms $s$-wave and $p$-wave contributions. We neglect all other higher order 
contributions in $\langle\sigma v\rangle$.
	 
	 \item Most importantly, in our prescribed methodology we assume the non-relativistic ($NR$) limit to compute and also expand the expression
	 for the thermally averaged cross-
section $\langle\sigma v\rangle$.
	 
	 \item In our analysis, we consider maximum mass of the dilaton dark matter to be ${\cal O }(1 ~ {\rm TeV})$. But our conclusions will remain unchanged for higher masses, as long as they satisfy the relic density constraint.
	 Higher the mass we consider, larger will be the scale of our effective theory.
	 
	\end{enumerate}

	The plan of the paper is as follows.
	
	\begin{itemize}
	 \item In section~\ref{s1}, we propose background models of extended theories of gravity, which are minimally coupled to SM fields. Initially we start with a model where the usual Einstein gravity is minimally 
	 coupled with the SM sector. But such a theory is not able to explain the genesis of dark matter at all. To explain this possibility without affecting the SM particle sector, we modify the gravity sector by allowing
	 quantum corrections motivated from (1) local $f(R)$ gravity and (2) non-minimally coupled dilaton with gravity and SM sector.
	 
	 \item In section~\ref{s3}, we construct our theory in the Einstein frame by applying conformal transformation on the metric. We explicitly discuss the rules and detailed techniques of conformal
	 transformation in the gravity sector as well as in the matter sector. For completeness, we present the results for arbitrary $D$ space-time dimensions. We use $D=4$ in the rest of our analysis. Then
	 we also show that the effective theory constructed from (1) local $f(R)$ gravity and (2) non-minimally coupled dilaton with gravity and SM sector looks exactly same.
	 Through conformal transformation, we derive the explicit form of dilaton effective potentials, which will be helpful to study the self interaction properties of the dark matter as well as the signatures of 
	 inflationary paradigm. In this paper, we have not explored this possibility. Detailed calculations are shown in section~\ref{s2} (Appendix A).
	 
	 \item In section~\ref{s4}, we use the relic constraint as observed by Planck 2015 to constrain the scale of the effective field theory $\Lambda_{UV}$ as well as the dark matter mass $M$. We consider two cases- (1) light
	 dark matter (LDM) and (2) heavy dark matter (HDM), and deduce upper bounds on thermally averaged cross section of dark matter annihilating to SM particles, in the non-relativistic limit. This classification
	 of DM into HDM and LDM is purely on the basis of the scale of the EFT considered. For LDM, the maximum mass of the DM candidate considered is less than ${\cal O }(350 {\rm GeV})$.
	 For HDM, DM masses between $350~{\rm GeV}$ and $1~TeV$ are considered. We shall find that for masses of DM greater than $350~{\rm GeV}$, the scale of the EFT increases by an order of magnitude, thereby leading to extra
	 suppression.
	 
	 \item In section~\ref{s10}, we explicitly discuss about the constraints on the parameters
	 of the background models of extended theories of gravity-
(1) local $f(R)$ gravity and (2) non-minimally coupled dilaton with gravity,
by applying the constraints from dark matter self interaction.
To describe this fact we consider the process $\phi\phi\rightarrow\phi\phi$, where $\phi$ is the
scalar dark matter candidate in Einstein frame as introduced earlier
by applying conformal transformation in the metric. Here $\phi\phi\rightarrow\phi\phi$ represents
dark matter self-interaction and characterized by the 
coefficient of $\phi^4$ term in the effective potential $V^{''''}_0$.
	  
	  \item In section~\ref{s5}, we propose different UV complete models from a particle physics point of view, which can give rise to the same effective theory that we have deduced from extended theories
	  of gravity. We mainly consider two models- (1) Inert Higgs Doublet model for LDM and (2) Inert Higgs Doublet model with a new heavy scalar for HDM. Thus, we have shown that UV completion of this effective theory need not come from
	  modifications to the matter sector, but rather from extensions of the gravity sector.

	 \item In section~\ref{s7}, we conclude with future prospects from this present work.

	\end{itemize}

	\begin{figure}[!h]
	      \centering
              \includegraphics[width=14cm,height=8.4cm]{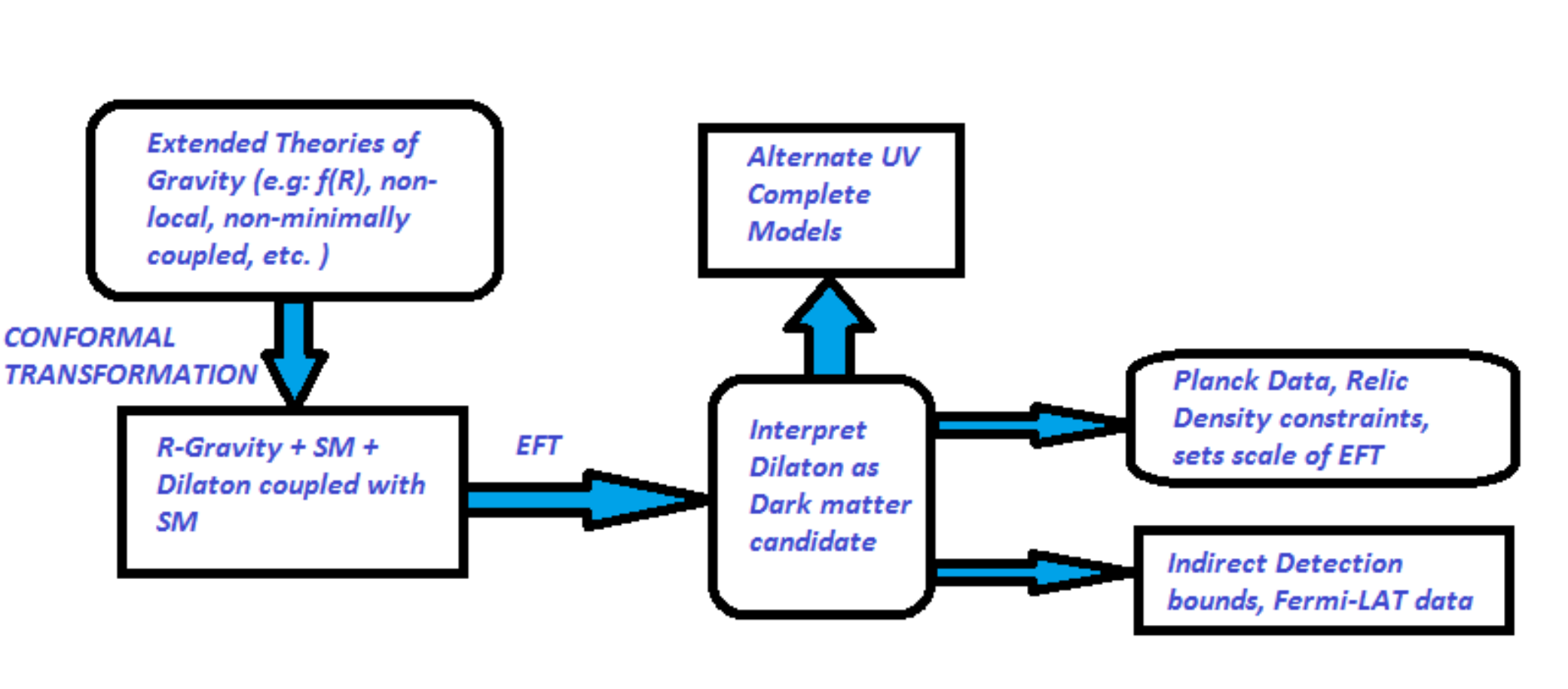}
              \caption{Schematic representation of the setup 
              which shows the complete algorithm of the described methodology in this paper.}
              \label{fig2}
           \end{figure}

%%%%%%%%%%%%%%%%%%%%%%%%%%%%%%%%%%%%%%%%%%%%%%%%%%%%%%%%%%%%%%%%%%%%%%%%%%%%%%%%%%%%
\section{The background  model}
\label{s1}
	In this section we start with the situation, where the well known Standard
	Model (SM) of particle physics in the matter sector is minimally coupled with
	the Einstein gravity sector and is described the following effective 
	action\cite{Sotiriou:2008rp}:
	\bea \label{eq1} S=\int d^{4}x\sqrt{-g}\left[\frac{\Lambda^2_{UV}}{2}R+{\cal L}_{SM}\right],\eea
	where $R$ is the Ricci scalar, ${\cal L}_{SM}$ is the SM Lagrangian density
	and $\Lambda_{UV}$ is the UV cut-off of the Effective Filed Theory as mentioned in the introduction of the paper~\footnote{The upper bound of the UV cut-off $\Lambda_{UV}$ is Planck scale $M_p$. }. 
	But it is important to mention here that, the effective action stated in Eq~(\ref{eq1})
	cannot explain the generation of a dark matter candidate without modifying the SM sector.
	
	To solve this problem, one needs to allow extensions in the standard Einstein gravity sector:
	\begin{enumerate}
		\item  By adding higher derivative and curvature terms in the effective action. For an
		example, within the framework
		of Effective Field Theory, one can incorporate local corrections in General Relativity (GR) in the gravity sector
		and write the action as \footnote{The Gauss-Bonnet gravity acts as a topological
			surface term in $D\leq 4$.}, \be S_{local}=\int d^{4}x
		\sqrt{-\!g}\left[
		\sum^{\infty}_{n=1}{\bf a}_{n}R^{n}+\sum^{\infty}_{m=1}{\bf b}_{m}\left(R_{\mu\nu}R^{\mu\nu}\right)^{m}
		+\sum^{\infty}_{p=1}{\bf c}_{p}\left(R_{\alpha\beta\delta\eta}R^{\alpha\beta\delta\eta}\right)^{p}\right].\ee  
		The co-efficients ${\bf a_{n}}, {\bf b_{m}}, {\bf c_{p}}$ of the correction factors 
		affects the ultraviolet behaviour of the gravity theory.
		But any arbitrary local modification of the renormalizable theory of GR typically contains
		massive ghosts which cannot be regularized using any standard field theoretic prescriptions.
		$f(R)$ gravity is one of the simplest versions of extended theory of gravity in which one fixes ${\bf a_{n}}\neq 0, {\bf b_{m}}=0$ and ${\bf c_{p}}=0$.
		Consequently, the effective action assumes the following simplified form:
		\be S_{local}=\int d^{4}x
		\sqrt{-\!g}f(R),\ee  
		where in general $f(R)$ is given by the following expression:
		\be f(R)=\sum^{\infty}_{n=1}{\bf a}_{n}R^{n},\ee
		which contains the full expansion in the gravity sector in terms of the Ricci scalar $R$. In principle, one can allow any combination of $f(R)$, but 
		to maintain renormalizability in the gravity sector, it is necessary to truncate the above infinite series in finite way. 
		String theory is one of the major sources through which it is possible to generate these types of corrections to the Einstein gravity sector by allowing quantum gravity effects.
		\item Considering non-minimal coupling between the Einstein gravity and additional scalar field, one can serve a similar purpose.
		Firstly, in the matter sector we incorporate the effects of quantum correction through the interaction between heavy and light sector 
		and then integrate out the heavy degrees of freedom from the Effective Field Theory picture. This finally  
		allows an expansion within the light sector, which can be written as:
		\be\begin{array}{llll}\label{bng}
			\displaystyle S_{M}[\phi,\Psi]=\int d^{4}x \sqrt{-^{(4)\!}g}
			\left[{\cal L}[\phi]+{\cal L}_{Heavy}[\Psi]+{\cal L}_{Int}[\phi,\Psi]\right]\\
			\displaystyle ~~\underrightarrow{Integrate~~ out~~\Psi~}~~~e^{i S_{M}[\phi]}=\int [{\cal D}\Psi] e^{iS_{M}[\phi,\Psi]}\\
			\displaystyle S_{M}[\phi]=\int d^{4}x \sqrt{-^{(4)\!}g}
			\left[{\cal L}[\phi]+\sum_{\alpha}J_{\Delta_{\alpha}}(g)\frac{{\cal O}_{\alpha}[\phi]}{\Lambda^{\Delta_{\alpha}-4}_{UV}}\right]
		\end{array}\ee
		where $J_{\alpha}(g)\forall \alpha $ are the Wilson coefficients which depend on the couplings $g$ of the full theory, and ${\cal O}_{\alpha}[\phi]$ are
		local operators having dimension $\Delta_{\alpha}$. All possible effective operators ${\cal O}_{\alpha}[\phi]$, which respect the symmetries of the
		full theory can be generated by this method. ${\cal L}[\phi]$ and ${\cal L}_{Heavy}[\Psi]$ describe the section which involves the light and heavy degrees of freedom, 
		and ${\cal L}_{int}[\phi,\Psi]$ consists of all
	        interactions amongst both sets of fields within Effective Field
		Theory prescription. After integrating out the
		heavy fields, the effective
		action has a renormalizable part:
		\be {\cal L}[\phi]=-\frac{g^{\mu\nu}}{2}(\partial_{\mu}\phi)(\partial_{\nu}\phi)-V_{ren}(\phi)\ee
		and a sum of non-renormalizable
		corrections denoted by ${\cal O}_{\alpha}[\phi]$, as given in Eq.~(\ref{bng}). Operators having dimensions less than four 
		are called ``relevant operators'' while those with dimensions greater than four are called ``irrelevant operators''.
		Theories having higher dimensional operators are dimensionally reduced to a four dimensional Effective Field Theory via various compactifications in string theory sector. However, corrections coming from graviton loops will 
		suppressed by the cut-off scale $\Lambda_{UV}$ which is fixed at Planck scale $M_p$, while those arising heavy sector will
		be suppressed by the background scale relevant for fields whose mass
		$M_{s}<\Lambda_{UV}\approx M_p$. Present observational status limits this scale around the GUT scale ($10^{16}$~{\rm GeV}). 
		In this context, we assume that the UV scale suppressed
	        operators will only modify the structure of the effective potential, without affecting the kinetic terms in the effective action.
		Consequently, these corrections will
		add with the renormalizable part of the potential $V_{ren}$  and give rise to the total potential $V(\phi)$ given by:
		\be V(\phi)=V_{ren}(\phi)+\sum^{\infty}_{\Delta_{\alpha}=5}J_{\Delta_{\alpha}}(g)\frac{\phi^{\Delta_{\alpha}}}{\Lambda^{\Delta_{\alpha}-4}_{UV}}=\sum^{\infty}_{\Delta_{\alpha}=0}C_{\Delta_{\alpha}}
		(g)\frac{\phi^{\Delta_{\alpha}}}{\Lambda^{\Delta_{\alpha}-4}_{UV}},\ee
		where $C_{\Delta_{\alpha}}$s are the Wilson coefficients. Thus the effective Lagrangian for the $\phi$ field is modified as:
		\be {\cal L}[\phi]=-\frac{g^{\mu\nu}}{2}(\partial_{\mu}\phi)(\partial_{\nu}\phi)-V(\phi).\ee
	\end{enumerate}
	Taking all these into account, the effective action for the background model can be expressed as:
	\be\begin{array}{lll}\label{eq3}
		\displaystyle   S=\left\{\begin{array}{lll}
			\displaystyle  
			\int d^{4}x\sqrt{-g}\left[\frac{\Lambda^2_{UV}}{2}f(R)+{\cal L}_{SM}\right]\,,~~~~~~ &
			\mbox{\small {\bf for {Case I }}}  \\ 
			%\displaystyle  
			%\int d^{4}x\sqrt{-g}\left[\frac{\Lambda^2_{UV}}{2}f(R,\Box)+{\cal L}_{SM}\right]\,,~~~~~~ &
			%\mbox{\small {\bf for {Case II }}}  \\ 
			\displaystyle   
			\int d^{4}x\sqrt{-g}\left[\frac{\Lambda^2_{UV}}{2}\left(1+\xi \frac{\phi^2}{\Lambda^2_{UV}}\right)R+{\cal L}[\phi]+{\cal L}_{SM}\right]\,~~~~~~ &
			\mbox{\small {\bf for {Case II}}}  
		\end{array}
		\right.
	\end{array}\ee
	where for {\bf Case I}, $f(R)$ represents any function of $R$ in general~\footnote{Technically
		only those functions of $R$ are allowed which gives rise to a renormalizable and ghost-free gravity theory.} and for {\bf Case II}, $\phi$ is the additional
	scalar field coupled to $R$ via non-minimal coupling $\xi$~\footnote{To avoid confusion, it is important to mention here that this possibility is completely different from the 
	situation where the SM Higgs field is coupled with the gravity sector via a non-minimal coupling.}.  Here for all three cases $\Lambda_{UV}$ represents the 
	Ultra-Violet (UV) cut-off scale for the Effective Field Theory. In this article,
	we will follow all possibilities with which we can study the effective theory of dark matter in detail.
	It is important to mention here that, all the effective actions are constructed in
	the {\it Jordan frame} of gravity. To explain the genesis of dark matter from
	the effective action, we have to apply {\it conformal transformation} in the metric,
	which transform the {\it Jordan frame} gravity to the {\it Einstein frame}. In the next section we
	discuss the technical details of conformal transformation in the extended gravity sector.
	
	\section{Construction of effective models from extended theories of gravity in Einstein frame}
	\label{s3}
	Conformal transformation of the metric is an appealing characteristic of the scalar-tensor theory of gravity \cite{Kanno:2002ia}
	which originates from superstring theory. Using this transformation, one can express the theory in
	two conformally related frames- Jordan and Einstein frames. In this paper, we use the {\it Einstein frame}
	to explain scalar dark matter generation in the context of Effective Field Theory. In the {\it Einstein frame}
	the new scalar field is coupled with the SM degrees of freedom via a conformal coupling factor. This new scalar field, {\it aka} ``scalaron'' or ``dilaton'', has a 
	geometrical origin and is generated 
	from the extended version of the gravity sector through conformal transformation in {\it Einstein frame}. In this section, we quote the results for dimension $D=4$, which will be used
	for further computation in the present context. The details of conformal transformation in arbitrary $D$ dimensions in explicitly computed in section~\ref{s2} (Appendix A).
	\subsection{Case I: From $f(R)$ gravity}
	\label{s3a}
	In case of $f(R)$ gravity, the conformal factor is given by:
	\be \label{eq8aa}\Omega(x)=e^{\omega(x)}=e^{\frac{1}{\sqrt{6}}\frac{\phi(x)}{\Lambda_{UV}}}=\sqrt{f^{'}(R)},\ee
	where $\phi$ is known as the ``scalaron'' or ``dilaton''.
	Here we start with the following action in {\it Jordan frame}:
	\be\begin{array}{lll}\label{eq12}
		\displaystyle   S= 
			\int d^{4}x\sqrt{-g}\left[\frac{\Lambda^2_{UV}}{2}f(R)+{\cal L}_{SM}\right]
	\end{array}\ee
	which can be recast in the following form:
	\be\begin{array}{lll}\label{eq13}
		\displaystyle   S= 
			\int d^{4}x\sqrt{-g}\left[\frac{\Lambda^2_{UV}}{2}f^{'}(R)R-U+{\cal L}_{SM}\right],
	\end{array}\ee
	where $U$ is defined as:
	\bea U=\frac{\Lambda^2_{UV}}{2}\left[f^{'}(R)R-f(R)\right].\eea
	Now transforming the {\it Jordan frame} action into {\it Einstein frame} we get finally:
	\bea S\Longrightarrow \tilde{S}=\int d^{4}x\sqrt{-\tilde{g}}\left[\frac{\Lambda^2_{UV}}{2}\tilde{R}
	-\frac{1}{2}\tilde{g}^{\mu\nu}\partial_{\mu}\phi\partial_{\nu}\phi-V(\phi)+e^{-\frac{2\sqrt{2}}{\sqrt{3}}\frac{\phi}{\Lambda_{UV}}}{\cal L}_{SM}\right], \eea
	where the effective potential in {\it Einstein frame}
	is given by:
	\be\label{eff} V(\phi)=\frac{U}{(f^{'}(R))^2}=\frac{\Lambda^2_{UV}}{2}\frac{f^{'}(R)R-f(R)}{(f^{'}(R))^2}.\ee
	For the further computation we will take the following structures of the function $f(R)$ as~\footnote{Here {\bf Case A1} and {\bf Case B1} represent Starobinsky model and scale free theory of gravity respectively.}:
	\be\begin{array}{lll}\label{eq14}
		\displaystyle   f(R)=aR+bR^{n}=\left\{\begin{array}{lll}
			\displaystyle  
			aR+bR^2\,,~~~~~~ &
			\mbox{\small {\bf with {Case A1: $a\neq 0, b\neq 0, n=2$ }}}  \\ 
			\displaystyle  
			bR^2\,,~~~~~~ &
			\mbox{\small {\bf with {Case B1: $a= 0, b\neq 0, n=2$ }}}  \\ 
			\displaystyle   
			aR+bR^{n}\,~~~~~~ &
			\mbox{\small {\bf with {Case C1: $a\neq 0, b\neq 0, n>2$ }}}  
		\end{array}
		\right.
	\end{array}\ee
	Now using Eq~(\ref{eq14}) in Eq~(\ref{eq8aa}) we get:
	\be\begin{array}{lll}\label{eq15}
		\displaystyle   \Omega(x)=e^{\omega(x)}=e^{\frac{1}{\sqrt{6}}\frac{\phi(x)}{\Lambda_{UV}}}=\left\{\begin{array}{lll}
			\displaystyle  
			\sqrt{\left(a+2bR\right)}\,,~~~~~~ &
			\mbox{\small {\bf with {Case A1: $a\neq 0, b\neq 0, n=2$ }}}  \\ 
			\displaystyle  
			\sqrt{2bR}\,,~~~~~~ &
			\mbox{\small {\bf with {Case B1: $a= 0, b\neq 0, n=2$ }}}  \\ 
			\displaystyle   
			\sqrt{a+nbR^{n-1}},\,~~~~~~ &
			\mbox{\small {\bf with {Case C1: $a\neq 0, b\neq 0, n>2$ }}}  
		\end{array}
		\right.
	\end{array}\ee
	where $R$ is the Ricci scalar in {\it Jordan frame}. 
	
	Further reverting Eq~(\ref{eq15}) as:
	\be\begin{array}{lll}\label{eq16}
		\displaystyle  R=\left\{\begin{array}{lll}
			\displaystyle  
			\frac{1}{2b}\left(e^{\frac{2}{\sqrt{6}}\frac{\phi(x)}{\Lambda_{UV}}}-a\right)\,,~~~~~~ &
			\mbox{\small {\bf with {Case A1: $a\neq 0, b\neq 0, n=2$ }}}  \\ 
			\displaystyle  
			\frac{1}{2b}e^{\frac{2}{\sqrt{6}}\frac{\phi(x)}{\Lambda_{UV}}}\,,~~~~~~ &
			\mbox{\small {\bf with {Case B1: $a= 0, b\neq 0, n=2$ }}}  \\ 
			\displaystyle   
			\left\{\frac{1}{nb}\left(e^{\frac{2}{\sqrt{6}}\frac{\phi(x)}{\Lambda_{UV}}}-a\right)\right\}^{\frac{1}{n-1}},\,~~~~~~ &
			\mbox{\small {\bf with {Case C1: $a\neq 0, b\neq 0, n>2$ }}}  
		\end{array}
		\right.
	\end{array}\ee
	and also using Eq~(\ref{eq16}) in Eq~(\ref{eff}), the effective potential can be expressed as:
	\be\begin{array}{lll}\label{eq17}
		\displaystyle   V(\phi)=\left\{\begin{array}{lll}
			\displaystyle  
			\frac{\Lambda^2_{UV}}{4b}e^{-\frac{4}{\sqrt{6}}\frac{\phi(x)}{\Lambda_{UV}}}\left(e^{\frac{2}{\sqrt{6}}\frac{\phi(x)}{\Lambda_{UV}}}-a\right)^2\,,~~~~~~ &
			\mbox{\small {\bf with {Case A1: $a\neq 0, b\neq 0, n=2$ }}}  \\ 
			\displaystyle  
			\frac{\Lambda^2_{UV}}{4b}\,,~~~~~~ &
			\mbox{\small {\bf with {Case B1: $a= 0, b\neq 0, n=2$ }}}  \\ 
			\displaystyle   
			\frac{\Lambda^2_{UV}b(n-1)}{(nb)^{\frac{n}{n-1}}}e^{-\frac{4}{\sqrt{6}}\frac{\phi(x)}{\Lambda_{UV}}}\left(e^{\frac{2}{\sqrt{6}}\frac{\phi(x)}{\Lambda_{UV}}}-a\right)^{\frac{n}{n-1}}.\,~~~~~~ &
			\mbox{\small {\bf with {Case C1: $a\neq 0, b\neq 0, n>2$ }}}  
		\end{array}
		\right.
	\end{array}\ee
	Here for {\bf Case A1} and {\bf Case C1}, the effective potential takes part in dark matter self interaction and for {\bf Case B1}, it mimics the role of a cosmological constant at late times~\footnote{This possibility 
	is not important for our present discussion as it has no minimum, which is necessarily required to stabilize the dark matter. In the context of dark energy this plays significant role at late times.}. 
	It is important to note that, from {\bf Case A1} and {\bf Case C1}, inflationary consequences 
	can also be studied in the present context. But in this article, we have not explored this possibility.
	In this Appendix \ref{s9} we discuss about the effective potential which can be used to model dark matter self interaction. Using the results of this section derived
from $f(R)$ gravity theory, we
further constrain the parameters $a$, $b$ and $n$.
	\subsection{Case II: From non-minimally coupled gravity}
	\label{s3c}
	In case of non-minimally coupled gravity the conformal factor is given by:
	\be \label{eq8cc}\Omega(x)=e^{\omega(x)}=e^{\frac{1}{\sqrt{6}}\frac{\phi(x)}{\Lambda_{UV}}}=\sqrt{\left(1+\xi \frac{\phi^2}{\Lambda^2_{UV}}\right)},\ee
	Here we start with the following action in {\it Jordan frame}:
	\be\begin{array}{lll}\label{eq24}
		\displaystyle   S= 
			\int d^{4}x\sqrt{-g}\left[\frac{\Lambda^2_{UV}}{2}\left(1+\xi \frac{\phi^2}{\Lambda^2_{UV}}\right)R-\frac{1}{2}g^{\mu\nu}\partial_{\mu}\phi\partial_{\nu}\phi-V(\phi)+{\cal L}_{SM}\right].
	\end{array}\ee
	Now transforming the {\it Jordan frame} action in {\it Einstein frame} we get finally:
	\bea S\Longrightarrow \tilde{S}=\int d^{4}x\sqrt{-\tilde{g}}\left[\frac{\Lambda^2_{UV}}{2}\tilde{R}
	-\frac{1}{2}\tilde{g}^{\mu\nu}\partial_{\mu}\tilde{\phi}\partial_{\nu}\tilde{\phi}-V(\tilde{\phi})+\frac{{\cal L}_{SM}}{\left(1+\xi \frac{\phi^2}{\Lambda^2_{UV}}\right)^2}\right], \eea
	where one can introduce a redefined field $\tilde{\phi}$ which can be written in terms of the scalar field $\phi$ as:
	\be d\tilde{\phi}=\frac{\sqrt{\left(1+\xi (1-6\xi)\frac{\phi^2}{\Lambda^2_{UV}}\right)}}{\left(1+\xi \frac{\phi^2}{\Lambda^2_{UV}}\right)}d\phi,\ee
	or equivalently one can write:
	\be\begin{array}{lll}\label{eq25}
		\displaystyle   \tilde{\phi}=\left\{\begin{array}{lll}\small
			\displaystyle  
			\sqrt{6}\Lambda_{UV}\left\{\tan^{-1}\left[\frac{\sqrt{6}\xi\frac{\phi}{\Lambda_{UV}}}{\left(1+\xi (1-6\xi)\frac{\phi^2}{\Lambda^2_{UV}}\right)}\right]
			\right.\\ \left.~~~~~~~~\displaystyle -\sqrt{1-\frac{1}{6\xi}}\sin^{-1}\left[\sqrt{\xi(6\xi-1)}\frac{\phi}{\Lambda_{UV}}\right]\right\}\,,~~~~~~ &
			\mbox{\small {\bf with {Case A3: $\xi\neq 1/6$ }}}  \\ 
			\displaystyle  
			\frac{\Lambda_{UV}}{\sqrt{6}}\sin^{-1}\left[\sqrt{6}\frac{\phi}{\Lambda_{UV}}\right]\,.~~~~~~ &
			\mbox{\small {\bf with {Case B3: $\xi= 1/6$ }}}  
		\end{array}
		\right.
	\end{array}\ee
	For the sake of simplicity $\xi\neq 1/6$ situation can also be studied in the two limiting physical situations as given by:
	\be\begin{array}{lll}\label{eq26}
		\displaystyle   \tilde{\phi}\approx\left\{\begin{array}{lll}\small
			\displaystyle  
			\phi\,,~~~~~~ &
			\mbox{\small {\bf with {Case A3a: $\xi\neq 1/6,\frac{\phi}{\Lambda_{UV}}<<\frac{1}{\xi}$ }}}  \\ 
			\displaystyle  
			\frac{\sqrt{6}}{2}\Lambda_{UV}\ln\left(1+\xi \frac{\phi^2}{\Lambda^2_{UV}}\right)\,.~~~~~~ &
			\mbox{\small {\bf with {Case A3b: $\xi\neq 1/6,\frac{\phi}{\Lambda_{UV}}>>\frac{1}{\xi}$ }}}  
		\end{array}
		\right.
	\end{array}\ee
	Now using Eq~(\ref{eq26}) in Eq~(\ref{eq8cc}) we get:
	\be\begin{array}{lll}\label{eq27}
		\displaystyle   \Omega(x)=\sqrt{\left(1+\xi \frac{\phi^2(\tilde{\phi})}{\Lambda^2_{UV}}\right)}\approx\left\{\begin{array}{lll}\small
			\displaystyle  
			1\,,~~~~~~ &
			\mbox{\small {\bf with {Case A3a: $\xi\neq 1/6,\frac{\phi}{\Lambda_{UV}}<<\frac{1}{\xi}$ }}}  \\ 
			\displaystyle  
			e^{\frac{\tilde{\phi}}{\sqrt{6}\Lambda_{UV}}}\,.~~~~~~ &
			\mbox{\small {\bf with {Case A3b: $\xi\neq 1/6,\frac{\phi}{\Lambda_{UV}}>>\frac{1}{\xi}$ }}}
		\end{array}
		\right.
	\end{array}\ee
	Consequently the most generalized version of the effective potential in {\it Einstein frame}
	can be expressed as:
	\be\begin{array}{lll}\label{eq28}
		\displaystyle  V(\tilde{\phi})=\frac{V(\phi(\tilde{\phi}))}{\left(1+\xi \frac{\phi^2(\tilde{\phi})}{\Lambda^2_{UV}}\right)^2}=\frac{1}{\left(1+\xi \frac{\phi^2(\tilde{\phi})}{\Lambda^2_{UV}}\right)^2}
	\sum^{\infty}_{\Delta_{\alpha}=0}C_{\Delta_{\alpha}}
		(g)\frac{\phi^{\Delta_{\alpha}}(\tilde{\phi})}{\Lambda^{\Delta_{\alpha}-4}_{UV}}\\ \displaystyle=\left\{\begin{array}{lll}\small
			\displaystyle  
			\sum^{\infty}_{\Delta_{\alpha}=0}C_{\Delta_{\alpha}}
		(g)\frac{\tilde{\phi}^{\Delta_{\alpha}}}{\Lambda^{\Delta_{\alpha}-4}_{UV}}\,,~~~~~~ &
			\mbox{\small {\bf with {Case A3a: $\xi\neq 1/6,\frac{\phi}{\Lambda_{UV}}<<\frac{1}{\xi}$}}}  \\ 
			\displaystyle  
			e^{-\frac{4\tilde{\phi}}{\sqrt{6}\Lambda_{UV}}}\sum^{\infty}_{\Delta_{\alpha}=0}C_{\Delta_{\alpha}}
		(g)\frac{\Lambda^{4}_{UV}}{\xi^{\frac{\Delta_{\alpha}}{2}}}\left(e^{\frac{2\tilde{\phi}}{\sqrt{6}\Lambda_{UV}}}-1\right)^{\frac{\Delta_{\alpha}}{2}}\,,~~~~~~ &
			\mbox{\small {\bf with {Case A3b: $\xi\neq 1/6,\frac{\phi}{\Lambda_{UV}}>>\frac{1}{\xi}$ }}} 
		\end{array}
		\right.
	\end{array}\ee
	Here for {\bf Case A3a} and {\bf Case A3b} both the effective potentials take part in self interaction. Inflationary consequences can be studied from {\bf Case A3a} and {\bf Case A3b}.
	It is important to mention here that for {\bf Case A3a} as the conformal factor $\Omega(x)\sim 1$, the dark matter do not couple to SM constituents. So for our discussion, only  {\bf Case A3b} is important.
	In this Appendix \ref{s9} we discuss about the effective potential construction necessarily required
	for dark matter self interaction. Using the results of this section derived
from non-minimally coupled gravity theory we
further constrain the non-minimal coupling parameter $\xi$.
\section{Construction of Effective Field Theory of dark matter}
\label{s4}
	  In this section, we explicitly argue that the dilaton field, which is generated via conformal transformation on the metric, can act as a viable dark matter
	  candidate. To start with, we consider the effective action which we have derived in Einstein frame through conformal transformation. We use an Effective Field Theory approach
	  to generate constraints on the scale of extended theories of gravity (as discussed in the previous section) from dark matter relic density constraints~\footnote{In our discussion the scale of the extended theories of gravity
	  sets the cut-off scale of the effective theory.}. We also compare the results obtained from annihilation of the dark matter (to SM particles) in our effective field theory model with current observational bound set by FermiLAT \cite{Ackermann:2015zua}.
	  Later on, we cite some well-known UV complete theories which can also give rise to the proposed effective theory.

%%%%%%%%%%%%%%%%%%%%%%%%%%%
\subsection{Construction of the model}
	  To start with, we consider the following general action obtained from transforming the {\it Jordan frame} action in {\it Einstein frame} as:
	  \bea S=\int d^{4}x\sqrt{-g}\left[\frac{\Lambda^2_{UV}}{2}R
	-\frac{1}{2}g^{\mu\nu}\partial_{\mu}\phi\partial_{\nu}\phi-V(\phi)+e^{-\frac{2\sqrt{2}}{\sqrt{3}}\frac{\phi}{\Lambda_{UV}}}{\cal L}_{SM}\right]. \eea
	  For the rest of the paper, for the sake of simplicity, we rescale the UV cut-off as: \be \Lambda_{UV}\rightarrow \frac{\sqrt{3}}{\sqrt{2}}\Lambda_{UV}.\ee 
	  The effective field theory action in {\it Einstein frame} consists of the following {\it three} components: 
	  \begin{enumerate} 
	   \item Einstein gravity sector ($R$),
	   \item Dynamics for the dilaton ($\phi$)~\footnote{In our discussion the effect of the dilaton effective potential ($V(\phi)$) is not studied explicitly.},
	   \item Modified matter sector which incorporates the interaction between SM fields and the dilaton ($\phi$).
	  \end{enumerate}
          Here our prime objective is to interpret this scalar field dilaton as a dark matter candidate. To show this explicitly, we impose a 
	  ${\cal Z}_2$ symmetry on top of our additional SM symmetries. Under this symmetry, all SM fields are even and $\phi$ is odd. This prevents terms involving decay of
	  $\phi$. Now assuming this scale of new physics is large enough, we can perform an expansion of the interaction term between dilaton and SM field contents i.e. $e^{-\frac{\phi}{\Lambda_{UV}}}{\cal L}_{SM}$ as:
	  \begin{equation}\label{frac}
	 e^{-\frac{\phi}{\Lambda_{UV}}}{\cal L}_{SM}\xrightarrow{{\cal Z}_2}\left\{1+ \frac{\phi^2}{2\Lambda_{UV}^2} + 
	 \underbrace{{\cal O}\left(\frac{\phi^4}{\Lambda_{UV}^4}\right)+\cdots}_{\bf All~suppressed~contributions}\right\}{\cal L}_{SM}\approx \left\{1+ \frac{\phi^2}{2\Lambda_{UV}^2} \right\}{\cal L}_{SM}.
	  \end{equation}
	  In Eq~(\ref{frac}), the odd terms vanish in the series expansion of $e^{-\frac{\phi}{\Lambda_{UV}}}$ because of the imposed ${\cal Z}_2$ symmetry.
	  
	  In our computation we only focus on the second term of the expansion as all higher order contributions are suppressed. 
	  This tells us that in the zeroth order of the expansion, we have the SM. However, because of the modification to the
	  gravity sector, we get higher order contribution in the next to leading order, which will produce all required interactions between dilaton and SM field contents.
	  
	  At this point, it is important to mention that the origin of the scalaron is purely geometric. It is a manifestation of the modified nature
	  of gravity. To use the well known results associated with Einstein gravity, we apply conformal transformation on the metric and generate the
	  scalaron in the Einstein frame. However, once we have transformed to the Einstein frame, and expanded the terms in the Lagrangian, we get an
	  effective theory of scalar dark matter, where DM couples {\it universally} to all SM particles. While an effective theory of scalar dark matter has been
	  widely studied in the literature, most of these involve non-universal coupling of DM to SM, i.e, each higher dimensional term comes with a different coupling constant. 
	  The novelty in our work is UV completing the well known scalar DM effective field theory
	  from a modified gravity perspective, and at the same time considering a universal coupling DM.

%%%%%%%%%%%%%%%%%%%%%%%%%%%%%
\subsection{Constraints from dark matter observation}
	
	  From the nature of the interaction terms, we see that in this effective theory, dark matter couples to all Standard Model particles {\it universally}. We can have 
	  $ 2\rightarrow 2 $ annihilation channels, as well as $ 2\rightarrow 3$ and $2\rightarrow 4$ ones respectively. However, the latter processes are suppressed (due to phase space) and 
	  are not considered in the calculation of the relic density bounds~\footnote{For completeness we suggest the readers to see ref.~\cite{Chen:2013gya} from which we 
          follow the computational strategy in the present context.}.
	  
	  For two dark matter particles of mass $M$ annihilating into particles of mass $m$ amd $m^{\prime}$, the thermally averaged annihilation cross-section in non-relativistic limit (NR) is given by:
	  \be\langle\sigma v\rangle_{NR}=\frac{1}{32\pi M^2}\sqrt{\frac{4M^2}{s}}\sqrt{\frac{M^2}{s-4M^2}}\sqrt{1-\frac{(m+m')^2}{4M^2}}\sqrt{1-\frac{(m-m')^2}{4M^2}}\Sigma(s;M,m,m',\Lambda_{UV}) ,\ee
	  where the symbol $\Sigma(s;M,m,m',\Lambda_{UV})$ can be expressed as: 
	  \be \label{sigma} \Sigma(s;M,m,m',\Lambda_{UV})=\int \frac{d\Omega}{4\pi}\langle|{\cal M}(M,\Lambda_{UV})|^2\rangle .\ee
          For our case, the processes which contribute to the annihilation process have same particle final states of mass m. So for our case 
          \be\langle\sigma v\rangle_{NR}=\frac{1}{32\pi M^2}\sqrt{\frac{4M^2}{s}}\sqrt{\frac{M^2}{s-4M^2}}\sqrt{1-\frac{4 m^2}{4M^2}}\int \frac{d\Omega}{4\pi}\langle|{\cal M}(M,\Lambda_{UV})|^2\rangle  .\ee
	  Here $\langle\sigma v\rangle_{NR}$ is obtained
	  by substituting \be s\rightarrow 4M^2+M^2v^2,\ee
	  where $s$ is the Mandelstam variable, $\langle|{\cal M}(M,\Lambda_{UV})|^2|\rangle$ is the thermally averaged invariant matrix amplitude squared, and $v$ is the velocity of dark matter ($v\approx 10^{-3}$). 
          This leads to the following series expanded form of the thermally averaged cross-section in non-relativistic limit:
	  \be \langle\sigma v\rangle_{NR}=\underbrace{a(\Lambda_{UV}, M)}_{\bf s-wave} +\underbrace{b(\Lambda_{UV}, M)v^2}_{\bf p-wave}+\cdots .\ee
	  We calculate he expression for $a(\Lambda_{UV}, M)$ and $b(\Lambda_{UV}, M)$ for all the processes given later, and the final results are given in the appendix.
           
          Since all these processes are of
	  higher order and represented by six dimensional operators, they will always be suppressed by power of $\Lambda_{UV}^2$. For eg., if we are looking
	  at a process which involves the annihilation of a pair of DM particles to a pair of photons via this higher dimensional operator, the expression
	  for $\langle\sigma v\rangle_{NR}$ will be given by
	  \bea
	    \langle\sigma v\rangle_{NR_{A_\mu A^\mu}}&=&\frac{4 M^2 \cos^4(\theta_W)}{\pi \Lambda^4_{UV}}+ \frac{2 M^2 \cos^4(\theta_W)}{\pi \Lambda^4_{UV}}v^2\nonumber \\
	    &=&a_{NR_{A_\mu A^\mu}}(\Lambda_{UV}, M) +b_{NR_{A_\mu A^\mu}}(\Lambda_{UV}, M)v^2
	  \eea
	  where $M$ is the mass of the DM candidate and $\theta_W$ is the Weinberg angle. We will get similar expressions for other processes, and the results are quoted in the appendix.
          All these processes will contribute to the relic density.  

          So from now we know that $a(\Lambda_{UV}, M)$ and $b(\Lambda_{UV}, M)$ are functions of the effective theory scale $\Lambda_{UV}$ and dark matter mass $M$. 
          Other parameter and masses that appear in the computaion of $a(\Lambda_{UV}, M)$ and $b(\Lambda_{UV}, M)$ are fixed quantities.  
          So we write them in a functional form, $a(\Lambda_{UV}, M)$  and $b(\Lambda_{UV}, M)$.
          We calculate the relic density of dark matter $\Omega_{\bf DM} h^2$ from the resulting $\langle\sigma v\rangle_{NR}$ in the present context.
	  The expression for $\Omega_{\bf DM} h^2$ is given by the standard result \cite{Chen:2013gya}
	  \begin{equation}\label{eq29}
	   \Omega_{\bf DM} h^2(\Lambda_{UV}, M)=\frac{2.08\times 10^9 x_F~\text{{\rm GeV}}^{-1}}{M_{Pl}\sqrt{g_*(x_F)}\left(a(\Lambda_{UV}, M) +3\frac{b(\Lambda_{UV}, M)}{x_F}\right)},
	  \end{equation}
	  where $M_{Pl}$ is the Planck mass, given by, $M_{Pl}\approx 10^{19}{\rm {\rm GeV}}$.
	  Here $x_F$ is a parameter which characterises the freeze-out temperature $(T_{F})$ of the dark matter, given by:
	  \begin{eqnarray}
	   x_F(\Lambda_{UV}, M) &=&\frac{M}{T_F} \\ \nonumber
                                &=&\ln\left(c(c+2)g\sqrt{\frac{45}{8}}\frac{M~M_{Pl}}{2\pi^3 \sqrt{g_*(x_F)}}\frac{(a(\Lambda_{UV}, M)+6\frac{b(\Lambda_{UV}, M)}{x_F})}{\sqrt{g_*(x_F)}}\right),
	  \end{eqnarray}
	 where  $g_*(x_F)=106$ (for SM) is the effective number of degrees of freedom at freeze-out and  $c$ is evaluated recursively from the 
        constraint \be c(c + 2) = 1. \ee Since the annihilation cross section $\langle\sigma v\rangle\propto a(\Lambda_{UV}, M)$ in the leading order, Eq~(\ref{eq29}) shows that the
        relic density is inversely proportional to the annihilation cross section of DM.

\begin{figure}
	   \includegraphics[width=0.3\textwidth]{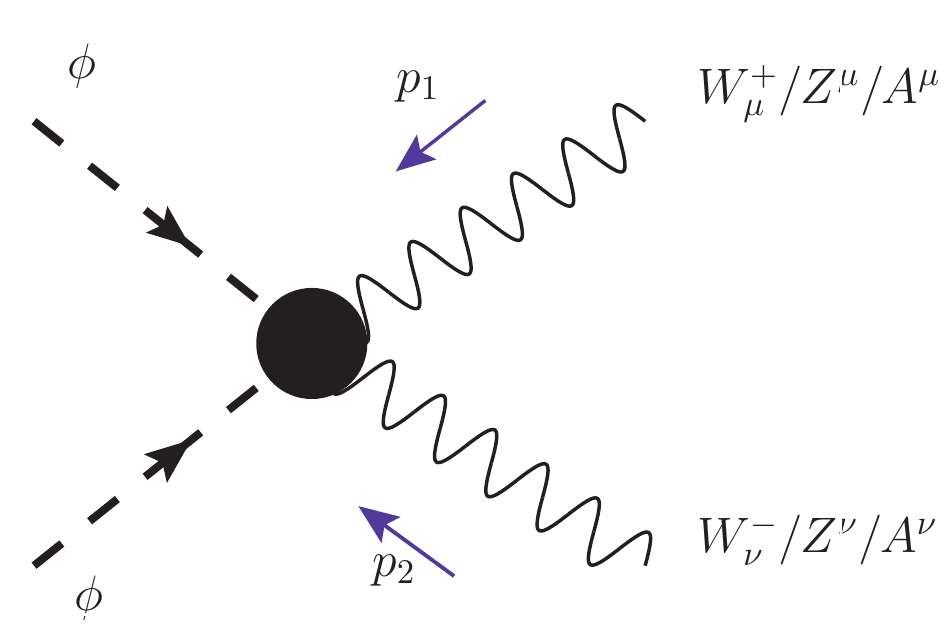}
           \includegraphics[width=0.3\textwidth]{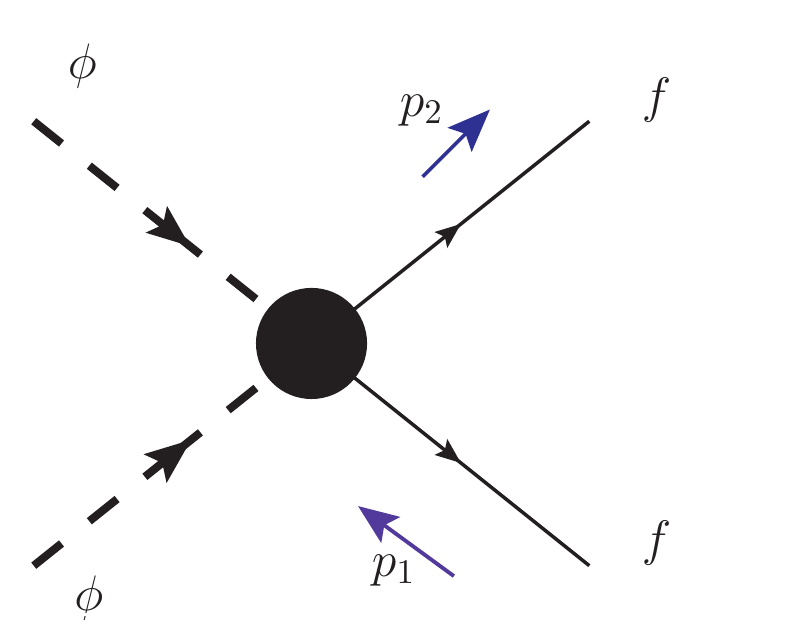}
           \includegraphics[width=0.3\textwidth]{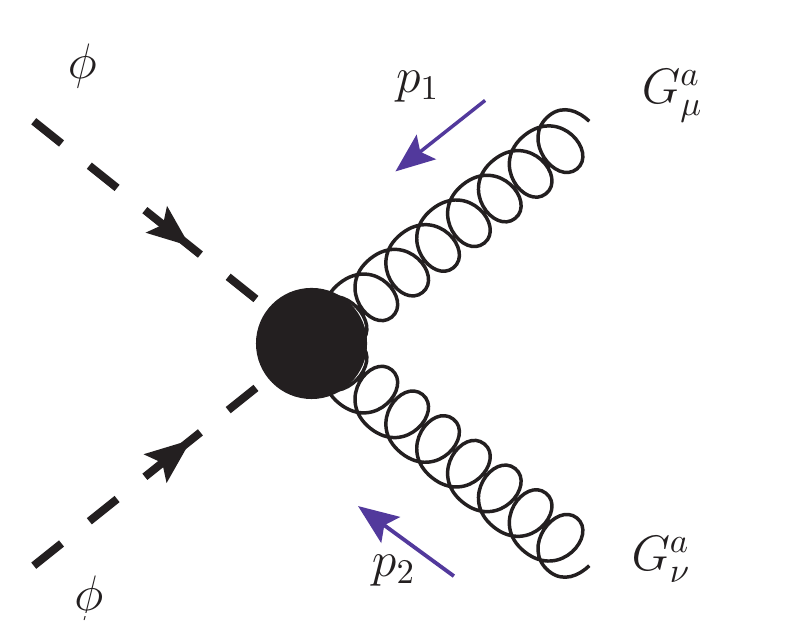}
           \includegraphics[width=0.3\textwidth]{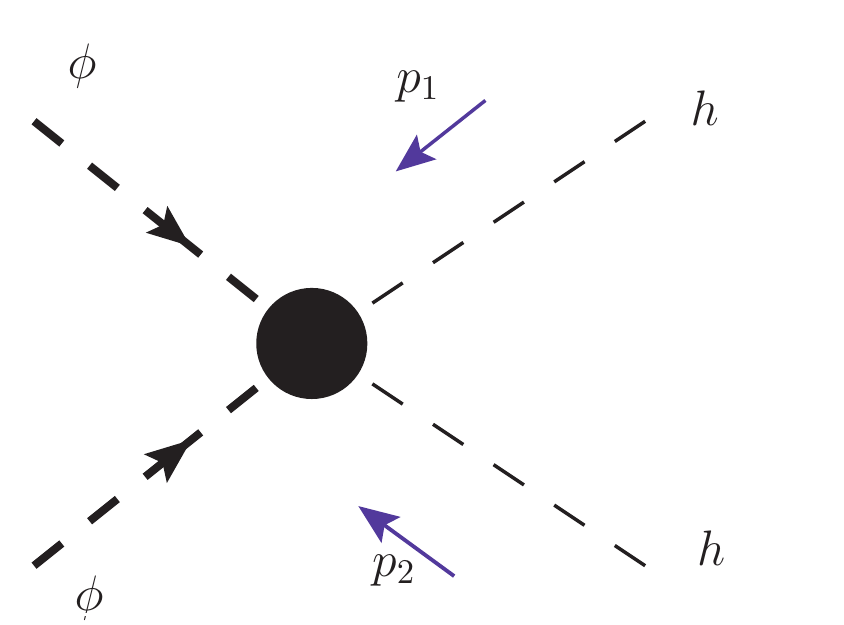}
           \includegraphics[width=0.3\textwidth]{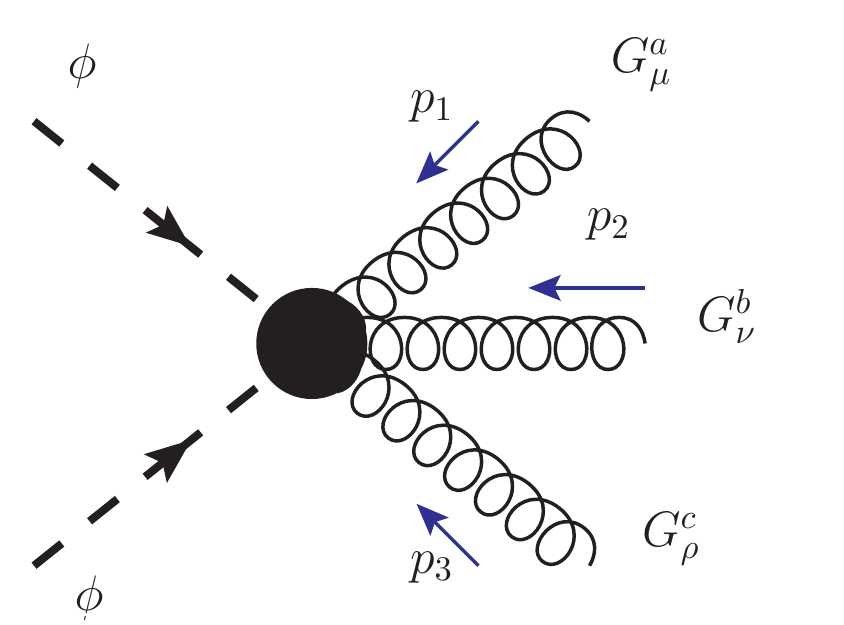}
           \includegraphics[width=0.3\textwidth]{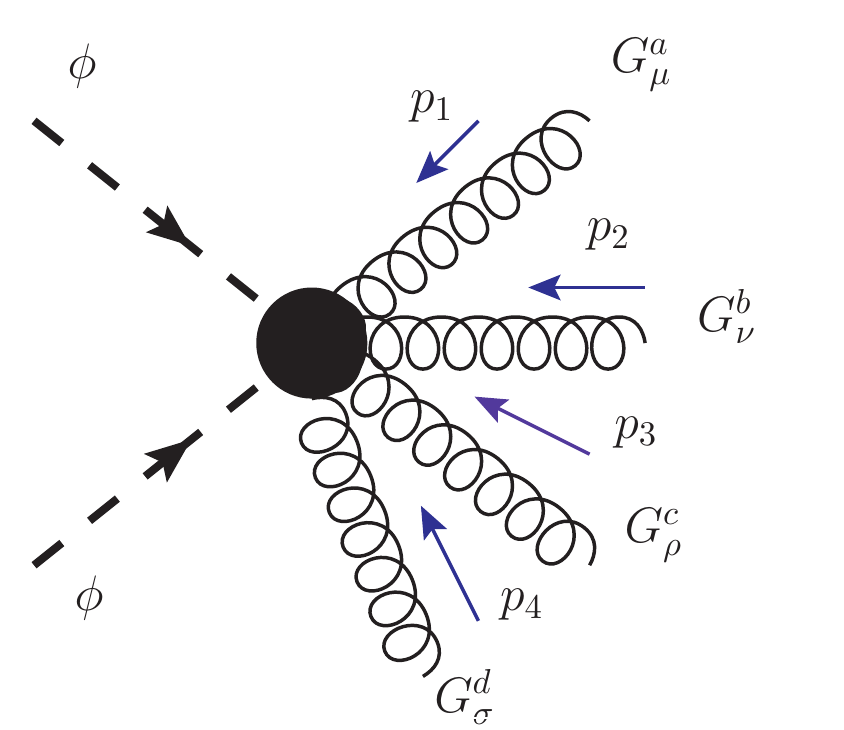}
           \includegraphics[width=0.3\textwidth]{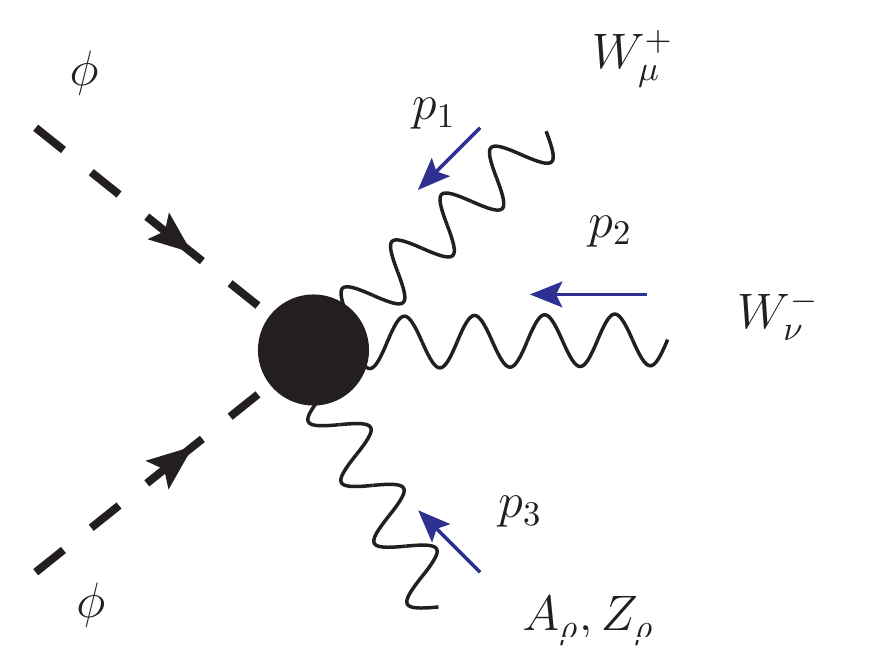}
           \includegraphics[width=0.3\textwidth]{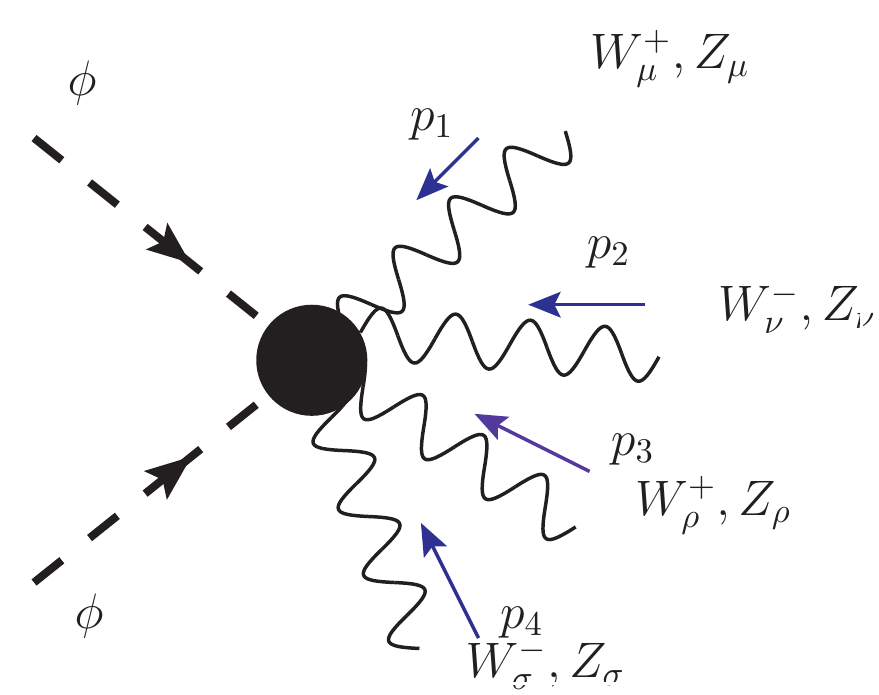}
           \includegraphics[width=0.3\textwidth]{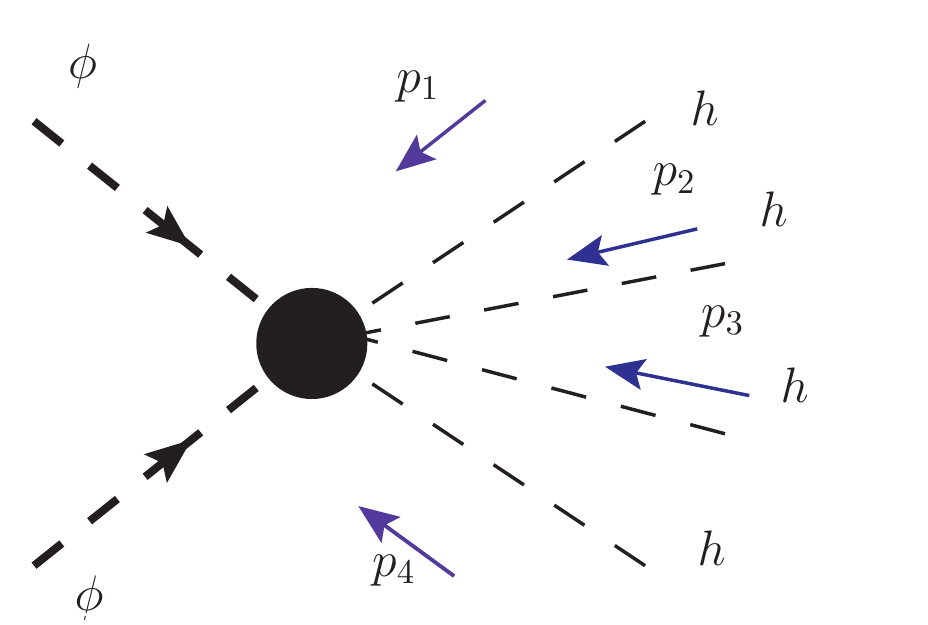}
           \caption{Effective processes contributing to relic density of dark matter. Here $2\rightarrow 3$ and $2\rightarrow 4$ processes are suppressed.}
           \label{fig1}
	  \end{figure}

	   In Eq~(\ref{eq29}), the unknown parameters are $\Lambda_{UV}$ and $M$. Therefore, demanding the value of $\Omega_{\bf DM} h^2$ to lie within the experimental bounds, we can get a range of $(M,\Lambda_{UV})$
	   satisfying the constraint obtained from recent Planck data \cite{Ade:2015xua}:
	   \be \Omega_{\bf DM} h^2 (\Lambda_{UV}, M)=0.1199\pm 0.0027.\ee Having obtained the relevant parameter space, we look at some of the well measured
	  annihilation channels for indirect detection of dark matter. These indirect detection experiments look for
	  dark matter annihilation to SM particles. We compare the results from our model
	  with the bounds given by FermiLAT \cite{Ackermann:2015zua} and others. The effective processes contributing to the relic density
	  calculation are shown in Fig~(\ref{fig1}). Keeping the above model in mind, in the next subsection we consider two possible scenarios:
	  \begin{enumerate}
	   \item Light Dark Matter (LDM),
	   \item  Heavy Dark Matter (HDM).
	  \end{enumerate} 
               The difference between the two scenarios is that, in case of HDM, the DM candidate has a mass greater than $350~{\rm GeV}$. In fig.~(\ref{fig2}), we have explicitly shown the allowed parameter space $(M,\Lambda_{UV})$ for our DM candidate. The plot shows visible 
               breaks at the mass of the top quark. It also shows that for masses of the DM candidate greater than $350~{\rm GeV}$ , the scales involved are larger by a factor of $10$. Thus, for HDM, processes 
               involving interactions with the DM will have an extra suppresion due to larger scales.
           This also imposes a constraint on the mass of the dilaton, if we are to interpret it as a DM candidate.

	   \begin{figure}[!h]
	      \centering
              \includegraphics[width=10cm,height=6.4cm]{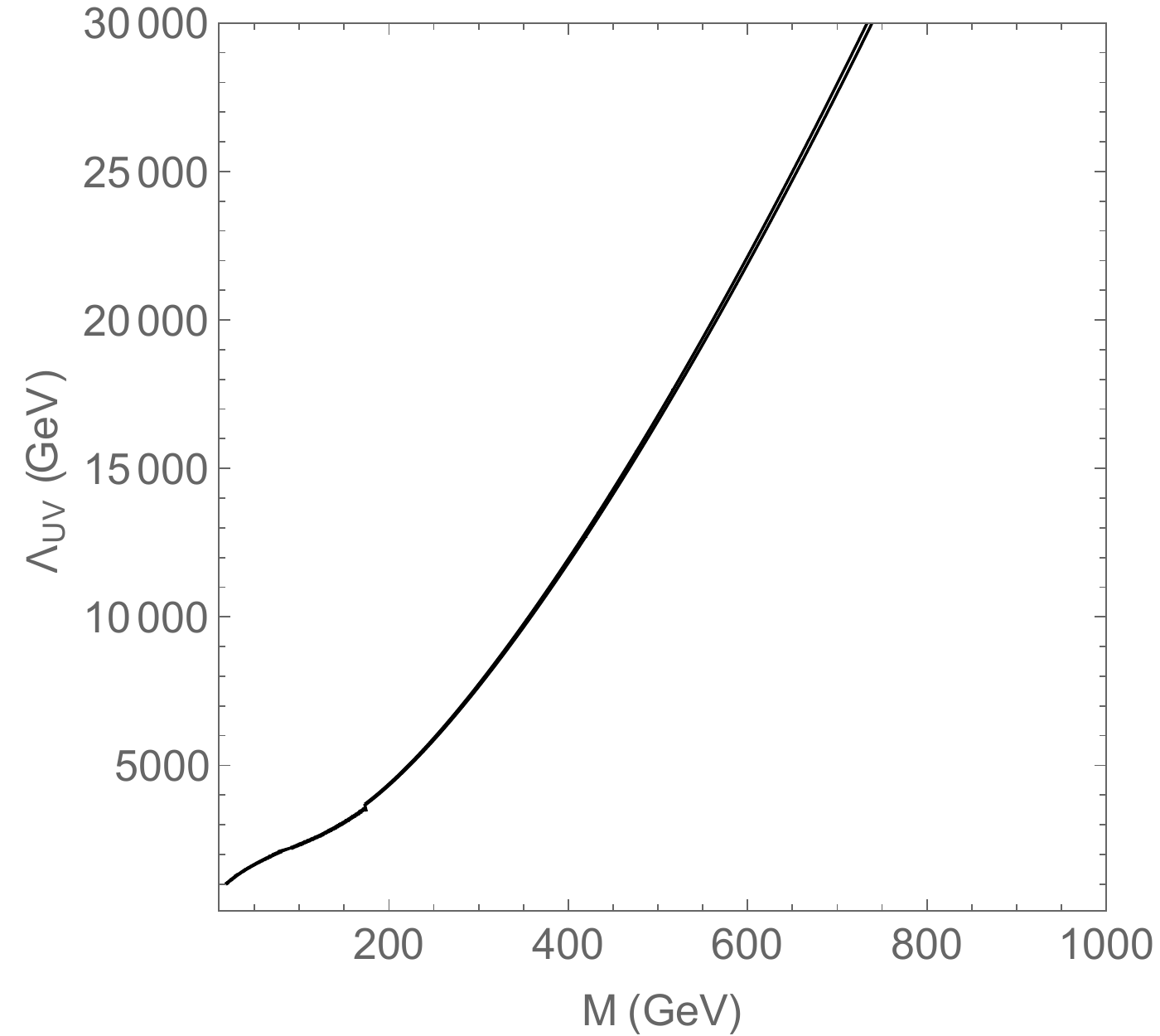}
              \caption{Allowed parameter space for the DM candidate.  The kink in the graph at $M=m_t$ shows that beyond this mass, annihilation channels to top pairs are allowed.}
              \label{fig2}
           \end{figure}

%%%%%%%%%%%%%%%%%	 
\subsubsection{Light Dark Matter}
\begin{figure}[!h]
\centering
\subfigure[$\langle\sigma v\rangle$ for LDM]{
    \includegraphics[width=10cm,height=5cm] {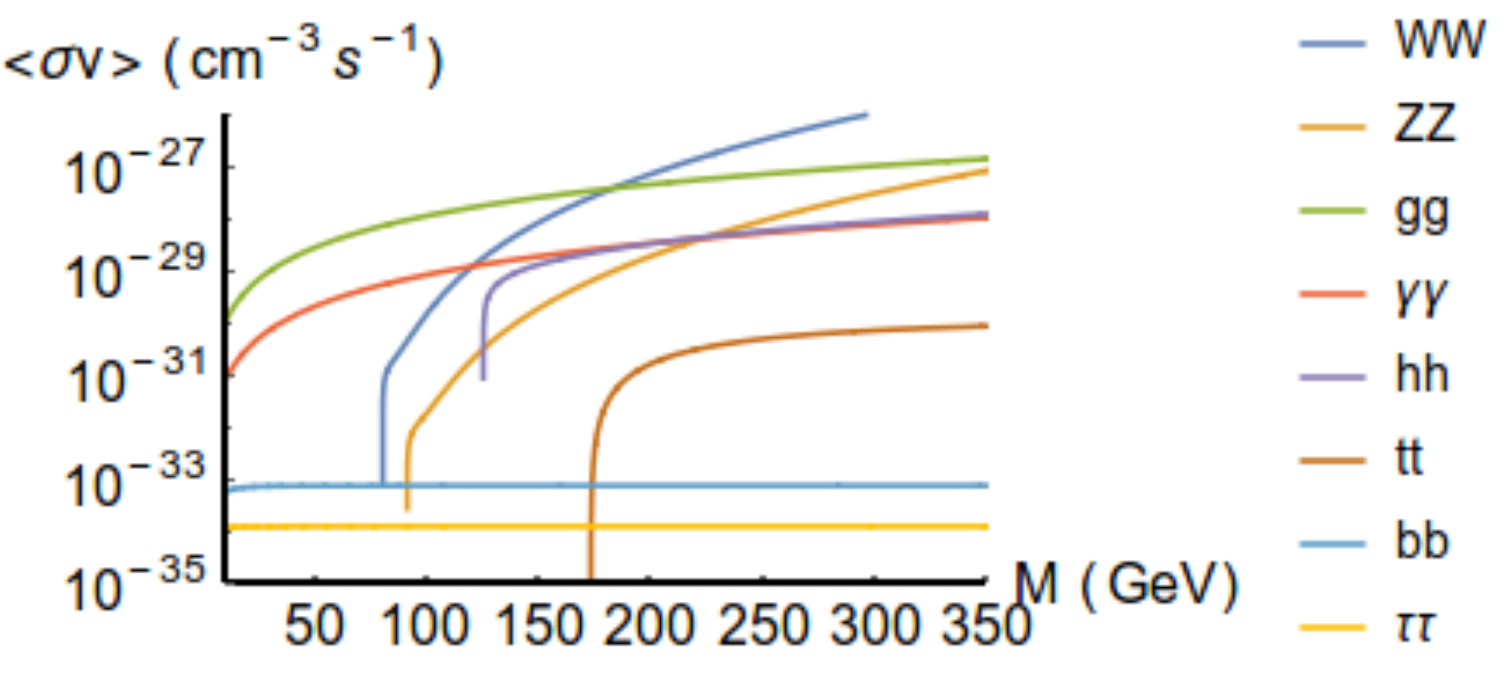}
    \label{x1}
    }
\subfigure[$\langle\sigma v\rangle$ for HDM]{
    \includegraphics[width=10cm,height=5cm] {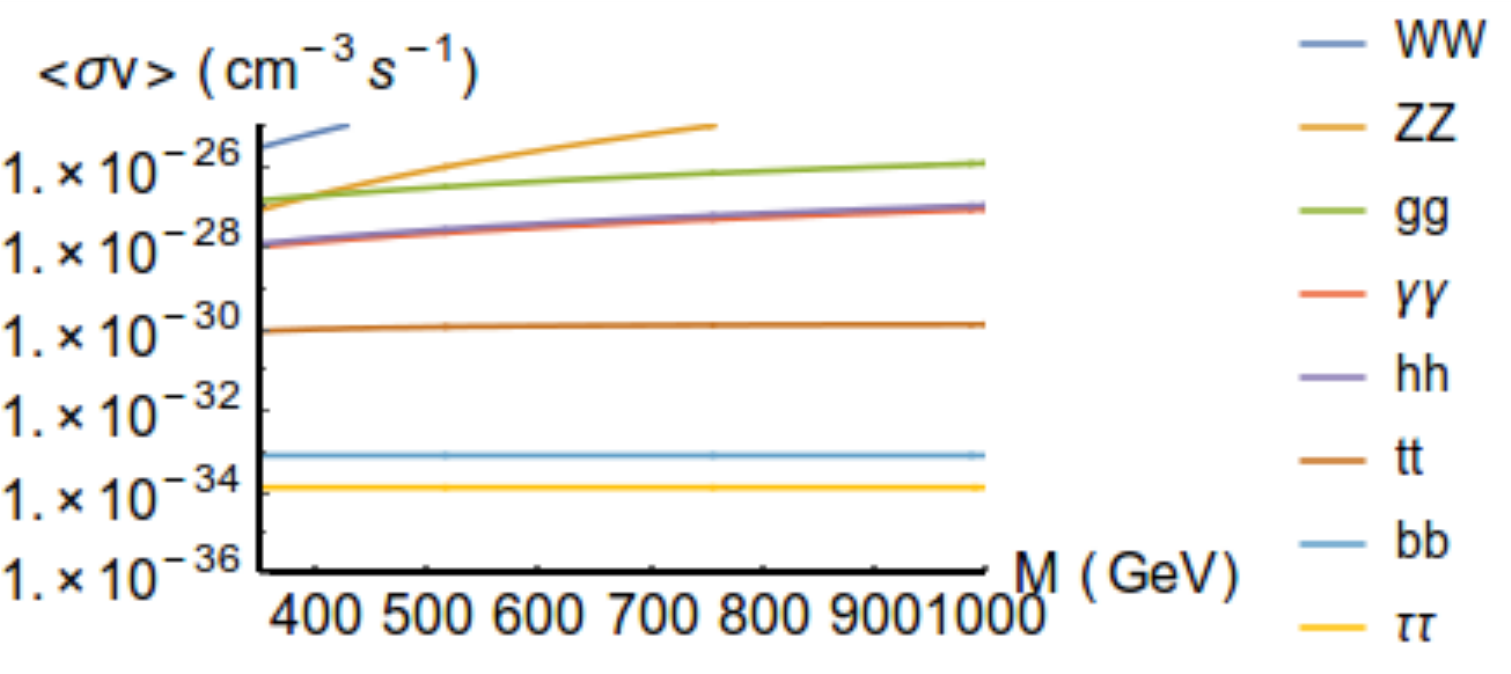}
    \label{x2}
    }
\caption[Optional caption for list of figures]{Top panel:Figure showing annihilation cross-sections of LDM to SM particles. Bottom panel: Figure showing annihilation cross-sections of HDM to SM particles.}
\label{fig3}
\end{figure}

	  In this subsection we consider that the dark matter candidate is a dilaton, with a mass less than  $350~ {\rm GeV}$. The main annihilation channels will be $DM~DM\rightarrow f~\bar{f}$ where
	   $f=t,b,\tau$, and $DM~DM\rightarrow gg, \gamma\gamma, W^+W^-,ZZ$ . Hence the total thermally averaged cross section for LDM can be written as:
	   \begin{equation}
	 \langle\sigma v\rangle_{LDM}= \langle\sigma v\rangle_{G_\mu G^\mu}+ \langle\sigma v\rangle_{A_\mu A^\mu} + \langle\sigma v\rangle_{W_\mu W^\mu}+ \langle\sigma v\rangle_{Z_\mu Z^\mu}+ \langle\sigma v\rangle_{hh}+\sum_{f} \langle\sigma v\rangle_{ff}
	   \end{equation}
	   In fig.~(\ref{x1}) , we show the allowed annihilation channels of LDM candidate into SM particles.

%%%%%%%%%%%%%%%%         
\subsubsection{Heavy Dark Matter}
         In this subsection we consider that the Dark Matter has a mass greater  $350~{\rm GeV}$. The annihilation channels remain the same, however as we can see from fig.(\ref{fig2}), the corresponding scale
         of the EFT increases by an order of magnitude.
 We also show the same annihilation channels as the LDM in fig.~(\ref{x2}).
          We observe similar features as observed in the previous case. However, the annihilation cross-sections are
           well below the current experimental sensitivity, and cannot be probed by present experiments. This extra suppression is mainly due to larger scales (by a factor of $10$)
           and universal ${ O(1)}$ coupling. 
 
 \begin{figure}[!h]
\centering
 \includegraphics[width=8cm,height=5cm] {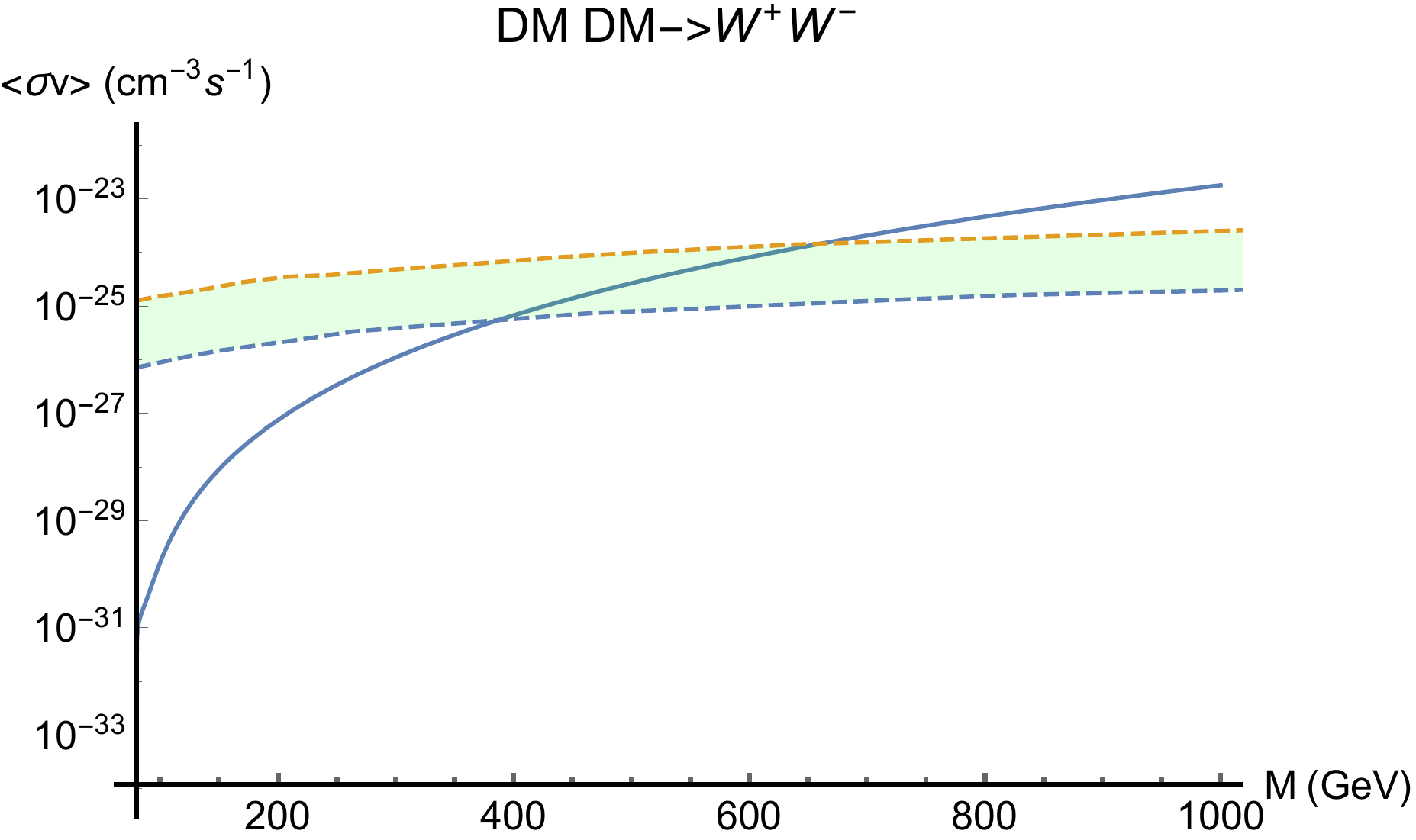}
\caption[Optional caption for list of figures]{Comparison of DM annihilation channel with bounds given by FermiLAT. }
\label{fig4}
\end{figure}
To show that these are well within the bounds given by FermiLAT ~\cite{Ackermann:2015zua}, we show one specific case
	   of DM annihilating into $W$ bosons in fig.~(\ref{fig4}) . The green shaded region shows $2\sigma$ bounds on the thermally averaged cross section for the process. We find that for most of our parameter space,
	   the predictions of our model are well within these bounds.

%%%%%%%%%%%%%%%%%%%%%%%%%%%%%

\section{Constraints from dark matter self interaction}
\label{s10}
In this subsection we will explicitly discuss about the constraints on the parameters of the background models of extended theories of gravity-
(1) local $f(R)$ gravity and (2) non-minimally coupled dilaton with gravity, by applying the constraints from dark matter self interaction.
To describe this fact let us consider the process $\phi\phi\rightarrow\phi\phi$, where $\phi$ is the scalar dark matter candidate in Einstein frame as introduced earlier
by applying conformal transformation in the metric. Here $\phi\phi\rightarrow\phi\phi$ represents dark matter self-interaction and characterized by the 
coefficient of $\phi^4$ term in the effective potential in Einstein frame i.e. estimated by the term $V^{''''}_0/4!$.

\begin{figure}
\centering
 \includegraphics[width=0.25\textwidth]{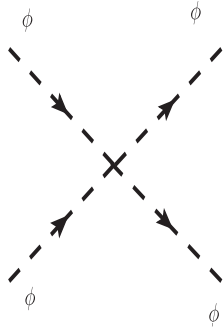}
 \caption{ DM-DM self interaction at the tree level}
 \label{self}
\end{figure}
The simplest four point contact interaction diagram contributing at the tree level is depicted in fig.~{\ref{self}}.
In this case the S-matrix element and amplitude of the $\phi\phi\rightarrow\phi\phi$ process is given by:
\bea i{\cal M}_{\phi\phi\rightarrow\phi\phi}&=&-i\lambda=-iV^{''''}_0/4!,\\ 
|{\cal M}_{\phi\phi\rightarrow\phi\phi}|^2&=&\lambda^2=\left(V^{''''}_0/4!\right)^2.\eea
Consequently the differential scattering cross section for the $\phi\phi\rightarrow\phi\phi$ process is given by:
\bea \frac{d\sigma}{d\Omega}&=& \frac{|{\cal M}_{\phi\phi\rightarrow\phi\phi}|^2}{64\pi^2 s}=\frac{\lambda^2}{64\pi^2 s}=\frac{\left(V^{''''}_0/4!\right)^2}{64\pi^2 s},\eea
where $s$ is the Mandelstum variable and in centre of mass frame characterized by $p_{1,2}=(M,0,0,\pm v)$ it is given by:
\bea \label{gh} s&=&(p_{1}+p_{2})^2=4M^2,\eea
where $p_{1,2}$ are the momenta of the two incoming scalar dark matter particle, $M$ is the mass of the scalar dark matter. Finally using Eq~(\ref{gh}) and integrating over the total solid angle 
one can finally write down the expression for the 
scattering cross section for the $\phi\phi\rightarrow\phi\phi$ self interaction process as:
\bea \label{gh2}\sigma&=& \frac{|{\cal M}_{\phi\phi\rightarrow\phi\phi}|^2}{64\pi M^2}=\frac{\lambda^2}{64\pi M^2}=\frac{\left(V^{''''}_0/4!\right)^2}{64\pi M^2}.\eea
Now, in order to have an observable effect on dark matter halos over large(cosmological) timescales, we have to
satisfy the following constraint in the present context~\cite{Kaplinghat:2013kqa}:
\bea\label{gh3} \frac{\sigma}{M}\lesssim 1~{\rm cm}^2/g=5\times 10^3 ~{\rm {\rm GeV}}^{-3}.\eea
Further using Eq~(\ref{gh2}) in Eq~(\ref{gh3}), we get the following simplified expression for this constraint: 
\bea\label{gh4} \frac{\lambda^2}{64\pi M^3}&\lesssim& 5\times 10^3 ~{\rm {\rm GeV}}^{-3},\nonumber\\
\Rightarrow\frac{\left(V^{''''}_0/4!\right)^2}{64\pi M^3}&\lesssim& 5\times 10^3 ~{\rm {\rm GeV}}^{-3}.\eea
Further depending on the different types of models of modified gravity theory as discussed in this paper, we will get a different value of the self-interaction parameter $\lambda$ ,
which is a function of some other parameters characterising the types of modified gravity. In our discussion for $f(R)$ gravity these parameters are $a$, $b$ and $n$, for non-minimally coupled dilaton with gravity and SM
it is characterised by the non-minimal coupling parameter $\xi$ as introduced earlier.
\subsection{Case I: For $f(R)$ gravity}
\underline{\bf A. For $n=2$:}\\
In this case $f(R)$ is fiven by:
\bea f(R)=a R+b R^2,\eea
where we set $a=1$ to have consistency with the Einstein gravity at the leading order and in this case $b$ is the only parameter that has to be constrained from dark matter self interaction .
Additionally it is important to note that the mass dimension of $b$ for $n=2$ case is $-2$.

In this case the self-interaction parameter $\lambda$ or $V^{''''}_0/4!$ can be expressed as:
\bea \label{gh6} \lambda&=& V^{''''}_0/4!={\cal M}_{\phi\phi\rightarrow\phi\phi}=\frac{14}{9b\Lambda^2_{UV}},\eea
where $\Lambda_{UV}$ is the UV cut-off of the effective field theory and further using Eq~(\ref{gh6}) the constraint condition stated in Eq~(\ref{gh4}) can be recast as:
\bea \label{gh5}|b|&\gtrsim& \frac{7}{1800\sqrt{2\pi}\Lambda^2_{UV}}\times \left(\frac{\rm {\rm GeV}}{M}\right)^{3/2}\\&=&\left\{\begin{array}{ll}
                    \displaystyle   3.87\times 10^{-13}~{\rm {\rm GeV}}^{-2} &
 \mbox{\small {\bf for ~LDM~with~$M=100~{\rm {\rm GeV}},~\Lambda_{UV}=2000~{\rm {\rm GeV}}$}}  \\ 
	\displaystyle 3.46\times 10^{-16}~{\rm {\rm GeV}}^{-2} &
 \mbox{\small {\bf for ~HDM~with~$M=500~{\rm {\rm GeV}},~\Lambda_{UV}=20000~{\rm {\rm GeV}}$}}.
 \end{array}
\right.\nonumber\eea
Thereby, depending on the choice of the dark matter mass $M$ and UV cut-off $\Lambda_{UV}$ parameters, we can constrain the parameter $b$ from dark matter self-interaction.\\
\underline{\bf B. For $n>2$:}\\
In this case $f(R)$ is fiven by:
\bea f(R)=a R+b R^n,~~~~~{\bf with~} n>2\eea
where for physical consistency, we set $a\neq 1$ and in this case, $a$ and $b$ are the parameters to be constrained from dark matter self interaction for $n>2$ case.
Here it is important to note that, for the further numerical estimation we set $n=3$. Additionally it is important to note that the mass dimension of $b$ for $n=3$ case is $-4$.

In this case the self-interaction parameter $\lambda$ can be expressed as:
\bea \label{gh61} \lambda&=& V^{''''}_0/4!={\cal M}_{\phi\phi\rightarrow\phi\phi},\eea
where $\Lambda_{UV}$ is the UV cut-off of the effective field theory.
Calculations give
\begin{equation}
 \lambda= \frac{0.0004+a\left[-0.0552+a\left(0.2405+(0.1140a-0.2958)a\right)\right]}{(1-a)^{2.5} b^{0.5} \Lambda^2_{UV}} \nonumber
\end{equation}

The allowed values of the parameters $a$ and $b$ for $n=3$ is shown in fig. \ref{x5}. This figure is shown for $M=100~ {\rm {\rm GeV}}$ and $\Lambda_{UV}=2~{\rm TeV}$.
The plot for the HDM candidate ($M=500~ {\rm {\rm GeV}}$ and $\Lambda_{UV}=20~{\rm TeV}$) look exactly the same. We observe that as $a$ approaches 1, the value of $b$ rises asymptotically and grows, whereas, for values of 
$a>1$, $b$ is negative and starts becoming smaller. We have checked that the nature of the results are similar for $n=4$ also, although the allowed values of $a$ and $b$ are slightly different.

\subsection{Case II: For non-minimally coupled gravity}
 Here we will discuss the situation where $\xi\neq 1/6,\frac{\phi}{\Lambda_{UV}}>>\frac{1}{\xi}$ and the effect of the non-minimal coupling $\xi$ can be visualized prominently as it 
 couples to the SM sector. The other case, $\xi\neq 1/6,\frac{\phi}{\Lambda_{UV}}<<\frac{1}{\xi}$, is not relevant in the present context as in this case the effect of the non-minimal coupling $\xi$ can be 
 neglected and SM sector couples to gravity minimally. In $\xi\neq 1/6,\frac{\phi}{\Lambda_{UV}}>>\frac{1}{\xi}$ case, the only parameter for the modified gravity theory is the non-minimal coupling $\xi$ for the 
 given value of dimensionless coefficients $C_{0}(g),C_{2}(g)$ and $C_{4}(g)$ .Here we 
 will constrain $\xi$ using the constraint from dark matter self interaction. For the sake of simplicity we set $C_{0}(g)\sim C_{2}(g)\sim C_{4}(g)\sim {\cal O}(1)$.
 
 In $\xi\neq 1/6,\frac{\phi}{\Lambda_{UV}}>>\frac{1}{\xi}$ the self-interaction parameter $\lambda$  can be expressed as:
\bea \label{gh62} \lambda&=&V^{''''}_0/4!={\cal M}_{\phi\phi\rightarrow\phi\phi}=\frac{14+\xi(16\xi-15)}{1944 \xi^2},\eea
where $\Lambda_{UV}$ is the UV cut-off the effective field theory. 

\begin{figure*}[!h]
\centering
\subfigure[$f(R)$ gravity, $n=3$]{
    \includegraphics[width=12cm,height=8.4cm] {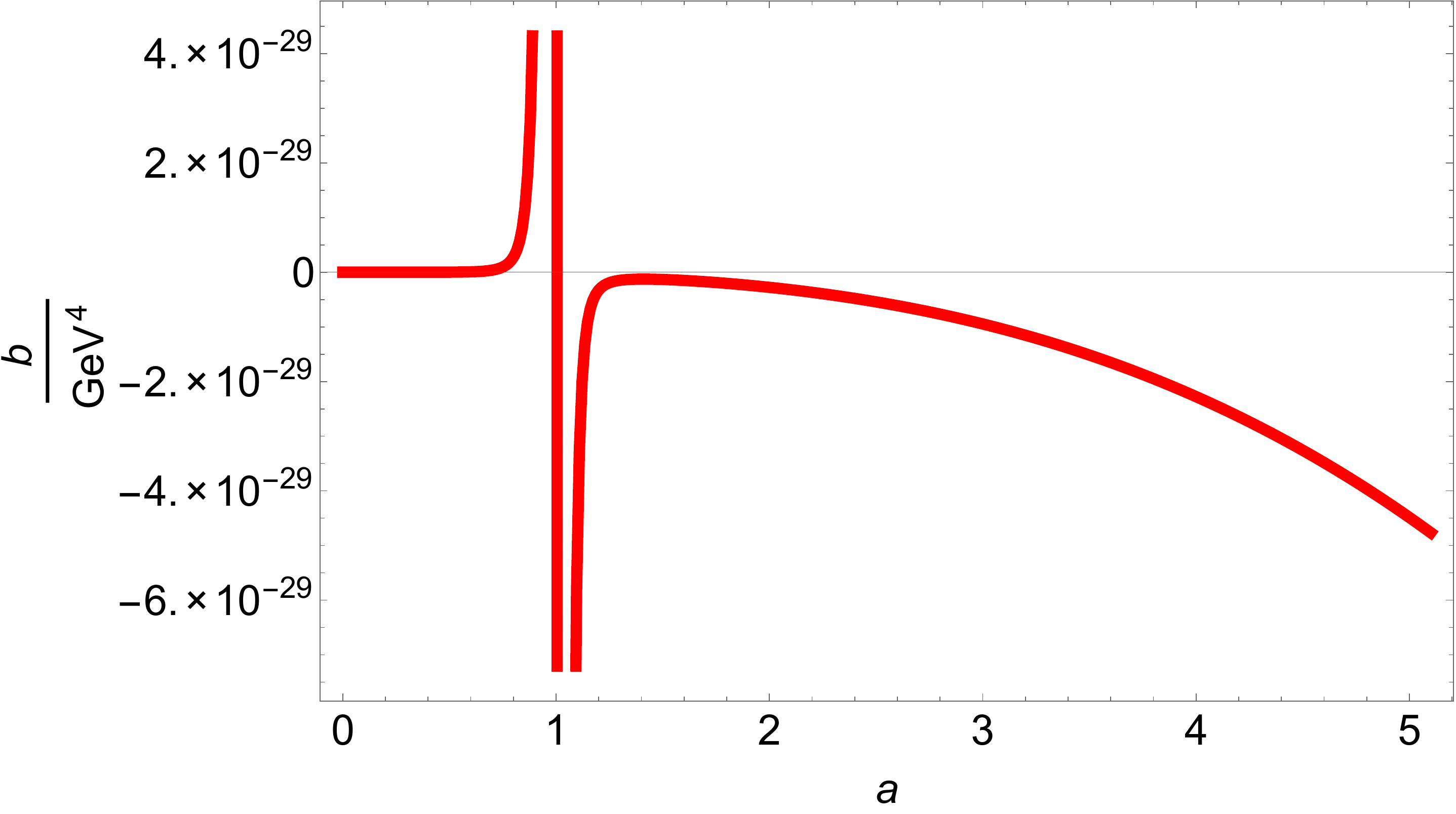}
    \label{x5}
}
\subfigure[Non Minimally Coupled Gravity]{
    \includegraphics[width=12cm,height=8.4cm] {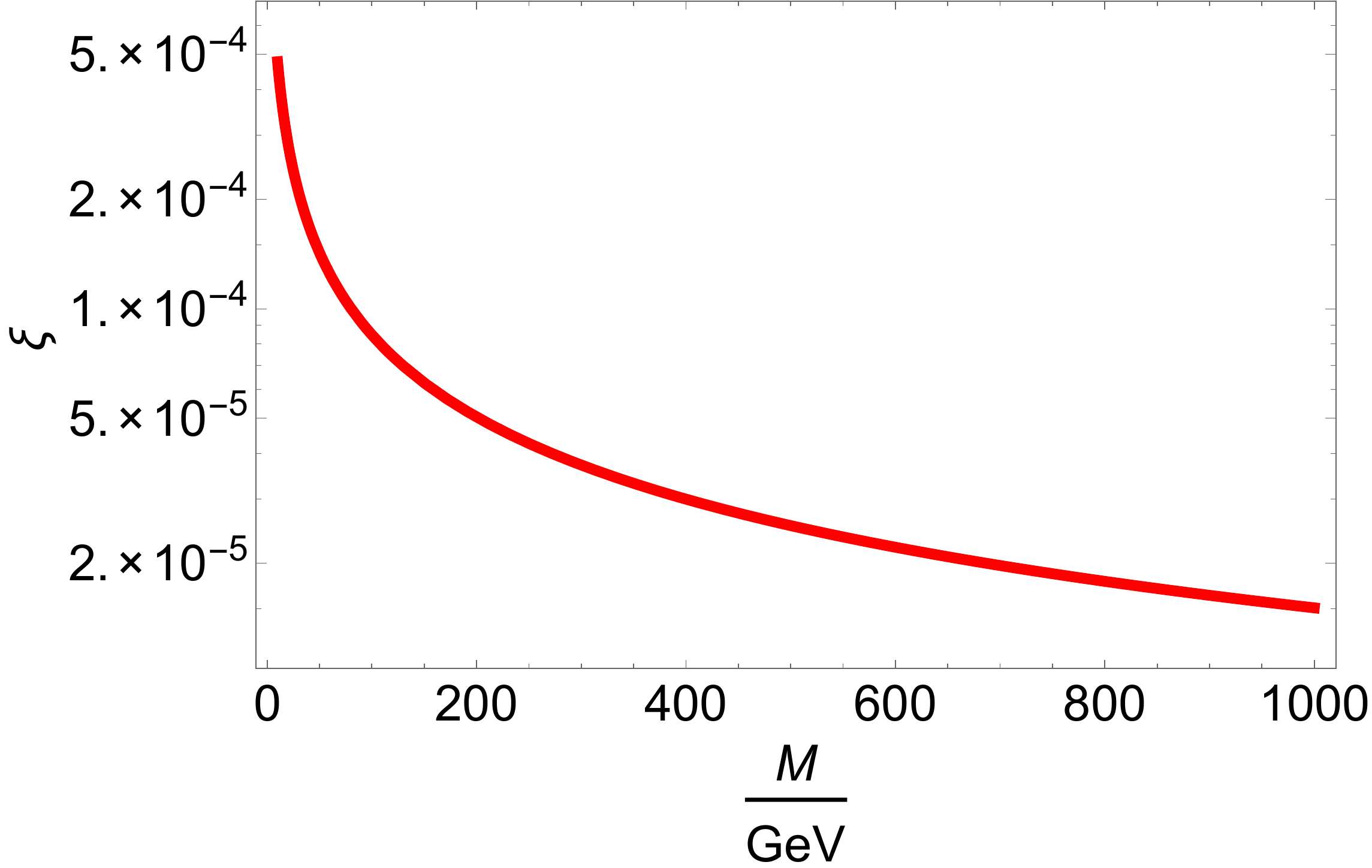}
    \label{x6}
}
\caption[Optional caption for list of figures]{Plots showing variation of the parameters of modified gravity. In fig. \ref{x5}, we show a variation of the 
parameters $a$ and $b$ in $f(R)=aR+bR^n$ for $n=3$. Notice the non-analytic behaviour at $a=1$. This graph is plotted for $M=100~ {\rm {\rm GeV}}$ and $\Lambda_{UV}=2~{\rm TeV}$.
The plots look exactly the same for the HDM candidate also.\\
In fig. \ref{x6}, we show a plot of the parameter $\xi$ of the non-minimally coupled gravity as a function of DM mass.}
\end{figure*}

In this case, we show a plot of the parameter $\xi$ as a function of $M$ in fig. \ref{x6}. We find that for a larger mass of the scalaron, a smaller value of $\xi\sim{\cal O}(10^{-5})$ is favored.
The range of $M$ is taken so as to cover the entire parameter space for LDM and HDM candidates. 

Thus, we observe that interpreting the dilaton as a dark matter candidate naturally incorporates dark matter self interaction and this can be directly used to put bounds on the parameters of the extended theories of 
gravity. We have presented a tree level analysis of the self interactions. This will receive corrections from higher order processes which have not been considered here.

%%%%%%%%%%%%%%%%%%%       
\section {Alternate UV completion of the Effective Field Theory}
\label{s5}
      In this section, we plan to highlight some of the well known models which behave similarly 
      as the effective field theory in the present context. Matter gravity interaction after a conformal transformation,
      generates terms involving interactions of the DM with other SM particles through the Lagrangian density,
\begin{align}
\mathcal{L}_{int} = \frac{\phi^2}{\Lambda_{UV}^2} \mathcal{L}_{SM},
\end{align} 
where $\Lambda_{UV}$ is the mass scale of the effective theory, below which this effective description works well.         

The usual procedure is to start with description of a UV complete theory. 
If the UV complete theory contains a heavy particle of mass ${\cal M} \sim \Lambda_{UV}$, we integrate out that 
particle to get an effective Wilsonian operator at energies less than the UV cut-off scale $\Lambda_{UV}$, 
which contains all other particles with masses lighter than $\Lambda_{UV}$. To compare one
UV complete model with the framework of effective description in the present context, we have to investigate if 
all the DM interaction operators are generated in that model. 

In order to quantify the validity of the effective field theory, we can compare its cross section with that from full theory at momentum transfer $Q_{tr}$ in the process,
\begin{align}
pp \rightarrow \phi \phi + jet / \gamma,
\end{align}
where $\phi$ is the scalar dark matter candidate in the model. The cross sections are calculated for $Q_{tr} < \Lambda_{UV}$, 
with $\Lambda_{UV}$ being the scale of the corresponding theory \cite{Busoni:2013lha,Busoni:2014sya,Busoni:2014haa}.
For the effective theory the scale can be taken arbitrarily but measurement of observables puts constraints on it. 
On the other hand, scale of a complete theory depends on particle to be integrated out from the theory. 
% In our case, we have effective theory in terms of $f(R)$ theory which provides us a $\Lambda$ in the theory. various constraints give possible order of magnitude of $\lambda$ values. We can check,
%comparing $f(R)$ theory with UV complete theories at those $\Lambda$, if our effective description is valid or not.

%%%%%%%%%%%%% 
\subsection{Inert Higgs Doublet Model for low $\Lambda_{UV}$}
Inert Higgs doublet model (IHDM) is a complete description where there is a DM 
candidate which can have interaction operators similar to the effective f(R) theory, at some particular mass scale. There are many studies in literature which look at the DM aspect of  
IHDM. A recent study\cite{Diaz:2015pyv} has treated the non-SM CP even scalar in the IHDM as the DM candidate and found out allowed parameter space satisfying the relic density. Part of this parameter 
space gets ruled out from the direct detection and collider physics constraints. An earlier study \cite{LopezHonorez:2006gr} analyses the DM relic abundance and 
prospects for direct or indirect detection in detail. Refs.\cite{LopezHonorez:2010tb,Honorez:2010re} discuss about new updated parameter regions in the IHDM. Ref.\cite{Gustafsson:2007pc} provides 
explanation of presence of $\gamma~$lines in the IHDM. 

\begin{figure}
\centering
{
    \includegraphics[width=1\textwidth] {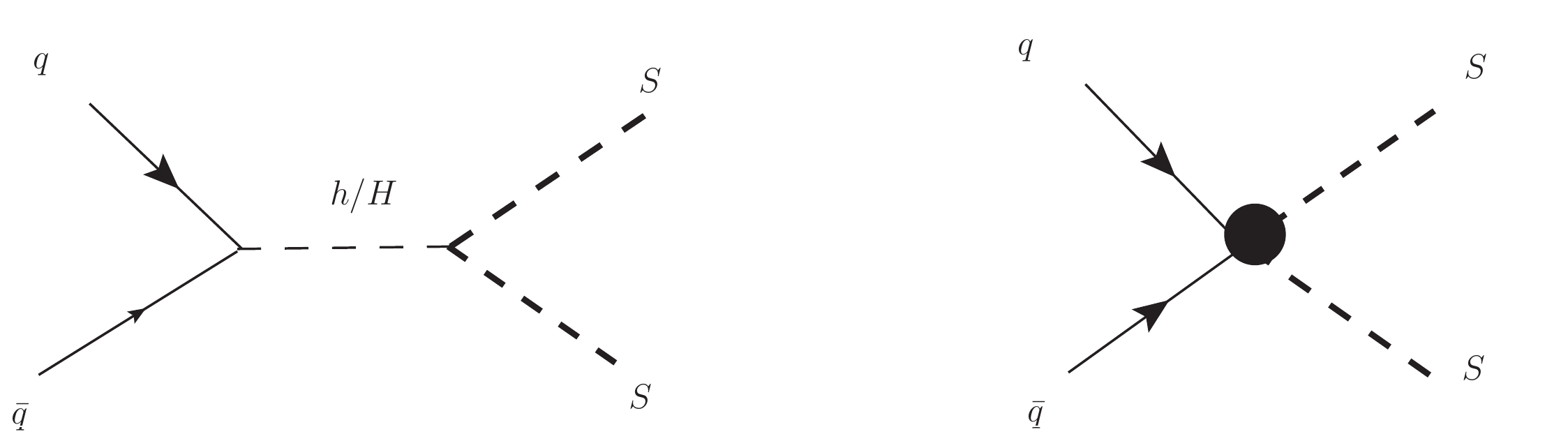}
    \caption{Left:$~q\bar{q}\rightarrow ~SS$ in the full theory. Right: Effective process, after integrating out the heavy mediator.}
    \label{UV}
}
\end{figure}

The Inert Higgs Doublet model is the minimal and simplest extension of the SM as it contains one extra scalar SU(2) doublet $\Phi_{2}$, apart 
from the SM-Higgs doublet $\Phi_{1}$ whose neutral component takes vacuum expectation value (vev) equal to v. 
It also couples to SM quarks and SM leptons similar to the SM-Higgs. $\Phi_2$ does not get any vev. It also does not couple to SM quarks and leptons. We also additionally enforce a 
$Z_2$ symmetry which transforms
 \bea 
\Phi_1 &\rightarrow & \Phi_1,\\
\Phi_2 &\rightarrow & -\Phi_2,
\eea
and other SM fields remain invariant under it.
Most general CP-invariant, $Z_2$ symmetry abiding scalar potential is given as:
\begin{align}
V (\Phi_1, \Phi_2) = \mu_{1}^{2} |\Phi_1|^{2} + \mu_{2}^{2} |\Phi_2|^{2} + \lambda_1 |\Phi_1|^{4} + \lambda_2 |\Phi_2|^{4} + \lambda_3 |\Phi_1|^{2} |\Phi_2|^{2} \nonumber \\+ \lambda_4 |\Phi_{1}^{\dagger} \Phi_2 |^{2} 
+ \frac{\lambda_5}{2} ((\Phi_{1}^{\dagger} \Phi_2)^2 + h.c. ),
\end{align}
where $\mu_{i}^{2}, \lambda_{i}~$s are taken real.
We define two scalar doublets in the unitary gauge as:
\bea 
\Phi_1 =  \bmat 0 \\ \frac{(v + h)}{\sqrt{2}} \emat  \ ; \qquad \Phi_2 =  \bmat H^{+} \\ \frac{( S + i A)}{\sqrt{2}} \emat .
\eea 
With these definitions we get the mass terms and the interaction Lagrangian of the scalar sector:
\begin{align}
\mathcal L \supset \frac{1}{2} m_h^2 h^2 + \frac{1}{2} m_S^2 S^2 + \frac{\lambda}{2} v h S S + \frac{\lambda}{4} h^2 S^2 + \frac{\lambda_2}{2} S^2 A^2 + \text{other interactions},
\end{align}
where \be  m_h^2= 2 \lambda_1 v^2, m_S^2 = \mu_2^2 + \frac{\lambda}{2} v^2 \ \ \text{with} \ \  \\ \lambda = \lambda_3 +\lambda_4 +\lambda_5,\ee 
and A is the CP-odd scalar of the model. Yukawa coupling in this theory is written as
\be  \mathcal L_{yuk} = y_q \bar{Q}_L \Phi_1 q_R + h.c., \ee 
which gives the mass of the fermions and also the $h \bar{q} q $ couplings. 
Due to the $Z_2$ symmetry imposed here, S can not decay to fermion channels. 
The $m_S$ can be made sufficiently small avoiding its decay to other scalars and WW/ZZ modes. 
Therefore we take $S$ as the DM candidate having direct interactions with the Higgs. 
This Lagrangian can give us processes like \be pp \rightarrow S S + jet/ \gamma\ee directly by a Higgs mediated process as shown in fig.~{\ref{UV}}. 
At $\Lambda_{UV} < m_h$, we can integrate out the Higgs boson to get effective vertex $~  \bar{q} q S S $, which is the kind of effective coupling to produce DM in the f(R) theory.
If we take S as the dilation then $f(R)$ theory in first order generates a coupling $\frac{m_{q}}{\Lambda_{UV}^2} \bar{q}q S S$.
In DM DM annihilation, processes with two final state particles contribute dominantly. We consider here the effective operators 
that only contribute to DM annihilation with two body final state. At $\Lambda_{UV} \sim m_h$ theory contains the DM candidate, W, Z boson and all SM fermions except the top quark. 
In IHDM heavy Higgs (h) has all SM like couplings i.e. standard Yukawa and hWW and hZZ couplings. 
Combining those with the hSS coupling present in the model we get effective operators of the form $\bar{q} q S S$, $W W S S$ and $Z Z S S$ integrating out the Higgs.
The couplings $h \gamma \gamma, hgg$ are present in the 1-loop level. So operators like $SS \gamma \gamma, SSgg$ also gets generated as the effective form of IHDM at $\Lambda_{UV} \sim m_h$. 
So we can generate all operators of $f(R)$ theory involving DM annihilation from the inert Higgs doublet model.
We can check the validity of the effective theory description of $f(R)$ gravity comparing it with the
inert 2HDM contributions to some process involving DM.

\subsection{UV complete model for high $\Lambda_{UV}$}
%When $\lambda_5$ is large in some parameter region of the IHDM, mass of the non-standard CP-even scalar (S) can be very large with CP-odd scalar (A) mass small.
%Couplings relevant to DM pair production process $pp \rightarrow A A + jet/ \gamma$ with S mediation are given,
%\begin{align}
%L \supset \frac{1}{2} m_S^2 S^2 + \lambda_{S}  S A A + y^{S}_{q} S \bar{q} q + ...,
%\end{align}
%where A is the DM candidate and $m_{S}$ is large. Here at $\Lambda_{UV} \sim m_S$ we can integrate out S from the theory to get effective vertex $\bar{q} q A A$, similar to f(R) theory effective operator 
%responsible for DM production.
We construct a model where we do not directly add effective operators contributing to DM pair production and DM annihilation processes as described above. We introduce a heavy scalar H as a part of third 
scalar doublet introduced in the IHDM. Here this new doublet acquires a non zero vev $v_H$, resulting in a non zero HAA/HSS vertex which originates from quartic coupling. Similarly H also couples to 
SM fermions and gauge bosons similarly as SM Higgs, though with different couplings. The Lagrangian consisting of H interaction terms is given as,
\begin{align}
L \supset \frac{1}{2} m_H^2 H^2 + \lambda_{H}  H S S + y^{H}_{q} H \bar{q} q + g^{H}_{V}H VV  +.. ,
\end{align}
where $V=\left\{W,Z\right\}$ and $q$ denotes any SM fermion. At $\Lambda_{UV} \sim m_H$, heavy scalar H gets integrated out from our model to provide effective operators like $\bar{q} q S S, VV S S$, which are similar to
the operators present in the effective f(R) theory. So with big $\Lambda_{UV}$ we can calculate DM cross sections.

	\section{Conclusion}
	\label{s7}
	To summarize, in the present article, we have addressed the following points:
	\begin{itemize}
	 \item In this paper, we have proposed background 
	models of extended theories of gravity, which are minimally coupled to SM fields. 
	Initially we have started with a model where the usual Einstein gravity is minimally 
	 coupled with the SM sector. But to explain the genesis of dark matter 
	 without affecting the SM particle sector, we have further modified the gravity sector by allowing
         quantum corrections motivated from (1) local $f(R)$ gravity
	 and (2) non-minimally coupled dilaton with gravity and SM sector.
	 
	 \item Next we have constructed an effective theory in the Einstein frame by applying conformal transformation on the metric.
	  We have explicitly discussed the rules and detailed techniques of conformal
	 transformation in the gravity sector as well as in the matter sector. Here for completeness, we have also presented the results
	 for arbitrary $D$ space-time dimensions. We have used $D=4$ in the rest of our analysis.
	 
	 \item Then
	 we have also shown that the effective theory constructed from (1) local $f(R)$ gravity and
	 (2) non-minimally coupled dilaton with gravity and SM sector looks exactly same.
	 
	 \item Here we have used the relic constraint as observed by Planck 2015 to constrain the scale of
	 the effective field theory $\Lambda_{UV}$ as well as the dark matter mass $M$. We have considered two cases- (1) light
	 dark matter (LDM) and (2) heavy dark matter (HDM), and deduced upper bounds on the thermally averaged cross
	 section of dark matter annihilating to SM particles, in the non-relativistic limit.
	 
	 \item We have modelled self-interactions of dark matter from their effective potentials in both cases-(1) local $f(R)$ gravity
	 and (2) non-minimally coupled dilaton with gravity and SM sector. Using the present constraint on dark matter self interactions, we have constrained the parameters of these two gravity models.
	 
	 \item Next we have proposed different UV complete models from a particle physics point of view, which
	 can give rise to the same effective theory that we have deduced from extended theories
	  of gravity. We have mainly considered two models- (1) Inert Higgs Doublet model for LDM and (2)
	  Inert Higgs Doublet model with a new heavy scalar for HDM. We have also explicitly shown that the UV completion of this effective field theory need not come from
	  modifications to the matter sector, but rather from extensions of the gravity sector.
	  
	  \item To conclude, we note that dark matter can indeed be considered to be an artifact of extended theories of gravity. In our work, we have presented a dark matter candidate which is generated purely from the
	  gravity sector. We have presented bounds on the mass of such a DM candidate, depending on the scale of the effective theory considered.

	\end{itemize}

		  The future prospects of this work are given below:
\begin{itemize}
 \item  The prescribed ideas can be worked out to derive cosmological constraints
for other modified gravity frameworks i.e. Randall Sundrum
single braneworld (RSII) \cite{Randall:1999vf,Maartens:2010ar,Brax:2004xh,Shiromizu:1999wj,Choudhury:2014sua,Choudhury:2011sq,Choudhury:2011rz,Choudhury:2012ib,Choudhury:2015jaa}
~\footnote{See also the refs.~\cite{Randall:1999ee,Choudhury:2013yg,Choudhury:2013eoa,Choudhury:2013aqa,Choudhury:2014hna,Choudhury:2015wfa}, 
for Randall Sundrum two braneworld (RSI) model.} ,Einstein-Hilbert-Gauss-Bonnet (EHGB) gravity \cite{Choudhury:2013eoa,Choudhury:2012yh,Choudhury:2015yna,Choudhury:2013dia},  Dvali-Gabadadze-Porrati
(DGP) braneworld \cite{Dvali:2000hr} and Einstein-Gauss-Bonnet-Dilaton (EGBD) gravity \cite{Choudhury:2013yg,Choudhury:2013aqa,Choudhury:2014hna,Choudhury:2015wfa,Choudhury:2016wlj} etc.
 
 \item Using the observational constraints from indirect detection of dark matter 
 one can further constrain various classes
of modified
theories of gravity scenario.
 
 %\item Another interesting feature of this model which has not been considered here is the effective potential term for the scalar dilaton $\phi$. This essentially signifies a self-interaction in the DM
 %sector, and can be used to model self-interacting dark matter\cite{Kaplinghat:2013kqa}. One can look for further possibilities in this area.
 
 \item Detailed study of DM collider and direct detection constraints \cite{Duerr:2015aka}
 on the effective theory prescription and the study of the effectiveness of the prescribed theory
from the various extended theories of gravity is one of the promising areas of research.

  \item Explaining the genesis of dark matter in presence of non-standard/ non-canonical
kinetic term \cite{Choudhury:2015hvr} and also exploring the highly non-linear regime of effective field theory
are open issues in this literature.

 \item The relation between dark matter abundance, primordial magnetic field and gravity waves and 
leptogenesis scenario from these effective operators can be studied. In
the case of RSII single membrane, some of the
issues have been recently worked out in ref.~\cite{Choudhury:2015eua}.
 
 \item The exact role of dark matter in the case of alternatives to inflation - specifically
for cyclic and bouncing cosmology \cite{Choudhury:2015baa,Choudhury:2015fzb,Choudhury:2015fzb} can also be studied in the present
context.
\end{itemize}

	%%%%%%%%%%%%%%%%%%%%%%%%%%%%%%%%%%%%%%%%%%%%%%%%%%%%%%%%%%%%%%%%%%%%%%%%%%%%%%%%%%%%%%%%%%%%%%%%%%%%%%%%%%%%%%%%%%%%%%%%%%%%%%%%%%%%%%%%%%%%%%%%%%%%%%%%%%%%%%%%%%%%%%%%%%%%%%%%%%%%%%%%%%%%%%%%%%%%%%%%%%%%%%%%%%%%%%%%%%%%%
	%%%%%%%%%%%%%%%%%%%%%%%%%%%%%%%%%%%%%%%%%%%%%%%%%%%%%%%%%%%%%%%%%%%%%%%%%%%%%%%%%%%%%%%%%%%%%%%%%%%%%%%%%%%%%%%%%%%%%%%%%%%%%%%%%%%%%%%%%%%%%%%%%%%%%%%%%%%%%%%%%%%%%%%%%%%%%%%%%%%%%%%%%%%%%%%%%%%%%%%%
	%\newpage
	\section*{Acknowledgments}
	SC would like to thank Department of Theoretical Physics, Tata Institute of Fundamental
Research, Mumbai for providing me Visiting (Post-Doctoral) Research Fellowship.
SC takes this opportunity to thank Sandip P. Trivedi, Shiraz
Minwalla, Soumitra SenGupta, Sayan Kar and
Supratik Pal for their constant support and inspiration. SC also thanks the organizers of School and Workshop on Large
Scale Structure: From Galaxies to Cosmic Web, The Inter-University Centre for Astronomy
and Astrophysics (IUCAA), Pune, India and specially Aseem Paranjape
and Varun Sahni for providing the academic visit during the work and giving the opportunity to present the 
work in the workshop. Also SC takes this opportunity to
thank the organizers of STRINGS 2015, International Centre for Theoretical Science,
Tata Institute of Fundamental Research (ICTS,TIFR) and Indian Institute of
Science (IISC), the organizers of National Strings Meet (NSM) 2015 and International
Conference in Gravitation and Cosmology (ICGC) 2015, Indian Institute of
Science Education and Research (IISER), Mohali for providing the local hospitality
during the work and giving a chance to discuss with Prof. Nima Arkani-Hamed. MS
would like to thank Amol Dighe, Basudeb Dasgupta and Disha Bhatia
for useful discussions and suggestions. Last but not the least, we would all like to
acknowledge our debt to the people of India for their generous and steady support
for research in natural sciences, especially for string theory, cosmology and particle
physics.

%%%%%%%%%%%%%%%%%%%%%%%%%
\section{Appendix A:~Conformal transformations in extended theories of gravity}
\label{s2}	

	\subsection{Conformal transformations in gravity sector}
	\label{s2a}
	Consider a $D$ dimensional space-time, where ${\cal M}$ is a smooth manifold and $g_{\mu\nu}$ is the Lorentzian metric on it.
	Under conformal transformation the metric $g_{\mu\nu}$, its inverse $g^{\mu\nu}$, determinant $\sqrt{-g}$ and the infinitesimal line element 
	transform as \cite{birell}:
	\bea\label{eq4a} g_{\mu\nu}\Longrightarrow\tilde{g}_{\mu\nu}&=&\Omega^{2}(x)g_{\mu\nu},\\
	\label{eq4b}g^{\mu\nu}\Longrightarrow\tilde{g}^{\mu\nu}&=&\Omega^{-2}(x)g^{\mu\nu},\\
	\label{eq4c}g_{\mu\nu}g^{\nu\beta}=\delta_{\mu}^{\beta}\Longrightarrow\tilde{g}_{\mu\nu}\tilde{g}^{\nu\beta}&=&\Omega^{2}(x)g_{\mu\nu}\Omega^{-2}(x)g^{\nu\beta}=\delta_{\mu}^{\beta},\\
	\label{eq4d}\sqrt{-g}\Longrightarrow \sqrt{-\tilde{g}}&=&\Omega^D(x)\sqrt{-g},\\
	\label{eq4e}ds^2=g_{\mu\nu}dx^{\mu}dx^{\nu}\Longrightarrow d\tilde{s}^2&=&\tilde{g}_{\mu\nu}d\tilde{x}^{\mu}d\tilde{x}^{\nu}=\Omega^{2}(x)g_{\mu\nu}dx^{\mu}dx^{\nu}=\Omega^{2}(x)ds^2\eea
	where the conformal factor $\Omega(x)$ is a smooth, non-vanishing, spacetime point dependent rescaling of the metric.
	The conformal transformations can shrink or stretch the distances
between the two points described by the same coordinate system $x^{\mu}$ (where $\mu=0,1,2,\cdots,D$) on the manifold ${\cal M}$.
However, these transformations preserve the angles between vectors, particularly null vectors, which define light
cones, thereby leading to a conservation of the global causal structure of the manifold. For simplicity, if
we take the conformal factor to be a constant space-time independent function, then it is known as a scale transformation.
On the contrary, any arbitrary $D$ dimensional coordinate transformations $x^{\mu}\rightarrow \tilde{x}^{\mu}$ only change the structural form of the 
coordinates, but not the associated geometry. This implies that coordinate transformations are completely different from conformal transformations, which connect two different frames via conformal couplings.

	Finally, the Einstein tensor transforms as:
	\bea\label{eq7v2} G_{\mu\nu}\Longrightarrow \tilde{G}_{\mu\nu}&=&G_{\mu\nu}+\left(\frac{D-2}{2}\right)\Omega^{-2}(x)\left[4\partial_{\mu}\Omega(x)\partial_{\nu}\Omega(x)
	+(D-5)\partial_{\alpha}\Omega(x)\partial^{\alpha}\Omega(x)g_{\mu\nu}\right]\nonumber\\
	&& ~~~~~~~~~~~~~~~~-(D-2)\Omega^{-1}(x)\left[\triangledown_{\mu}\triangledown_{\nu}-g_{\mu\nu}\Box\right]\Omega(x),\\ 
	\label{eq7v3} \tilde{G}_{\mu\nu}\Longrightarrow {G}_{\mu\nu}&=&\tilde{G}_{\mu\nu}+\left(\frac{D-2}{2}\right)\Omega^{-2}(x)(D-1)\partial_{\alpha}\Omega(x)\partial^{\alpha}\Omega(x)\tilde{g}_{\mu\nu}\nonumber\\
	&& ~~~~~~~~~~~~~~~~+(D-2)\Omega^{-1}(x)\left[\tilde{\triangledown}_{\mu}\tilde{\triangledown}_{\nu}-\tilde{g}_{\mu\nu}\widetilde{\Box}\right]\Omega(x).\eea
	We observe that, conformal transformations under some specific conditions behave like duality transformation in superstring theory. To demonstrate this, let us define the conformal factor as:
	\be \label{eq8}\Omega(x)=e^{\omega(x)}=e^{\frac{\kappa}{\sqrt{6}}\phi(x)},\ee
	where $\omega(x)=\frac{\kappa}{\sqrt{6}}\phi(x)$ represents the new scalar field ``scalaron'' or ``dilaton''. Here we define $\kappa=\Lambda^{-1}_{UV}$. Now, the conformal transformation in the metric $g_{\mu\nu}$, its inverse $g^{\mu\nu}$, determinant $\sqrt{-g}$ and consequently the infinitesimal line element 
	transform as:
	\bea\label{eq4aa} g_{\mu\nu}\Longrightarrow\tilde{g}_{\mu\nu}&=&e^{2\omega(x)}g_{\mu\nu}=e^{\frac{2\kappa}{\sqrt{6}}\phi(x)}g_{\mu\nu},\\
	\label{eq4bb}g^{\mu\nu}\Longrightarrow\tilde{g}^{\mu\nu}&=&e^{-2\omega(x)}g^{\mu\nu}=e^{-\frac{2\kappa}{\sqrt{6}}\phi(x)}g_{\mu\nu},\\
	\label{eq4cc}g_{\mu\nu}g^{\nu\beta}=\delta_{\mu}^{\beta}\Longrightarrow\tilde{g}_{\mu\nu}\tilde{g}^{\nu\beta}&=&e^{2\omega(x)}g_{\mu\nu}e^{-2\omega(x)}g^{\nu\beta}=\delta_{\mu}^{\beta},\\
	\label{eq4dd}\sqrt{-g}\Longrightarrow \sqrt{-\tilde{g}}&=&e^{D\omega(x)}\sqrt{-g}=e^{\frac{D\kappa}{\sqrt{6}}\phi(x)}\sqrt{-g},\\
	\label{eq4ee}ds^2=g_{\mu\nu}dx^{\mu}dx^{\nu}\Longrightarrow d\tilde{s}^2&=&\tilde{g}_{\mu\nu}d\tilde{x}^{\mu}d\tilde{x}^{\nu}=e^{2\omega(x)}g_{\mu\nu}dx^{\mu}dx^{\nu}=e^{\frac{2\kappa}{\sqrt{6}}\phi(x)}ds^2.\eea
	In the present context, the {\it Einstein frame} and the {\it Jordan frame} are connected via the following duality transformation:
	\be \Omega(x)=e^{\omega(x)}=e^{\frac{\kappa}{\sqrt{6}}\phi(x)}\Longleftrightarrow\Omega^{-1}(x)=e^{-\omega(x)}=e^{-\frac{\kappa}{\sqrt{6}}\phi(x)},\ee
	which is exactly same as the weak-strong coupling duality in superstring theory. Using Eq~(\ref{eq8}) we get:
	\bea \Omega^{-1}(x)\partial_{\mu}\Omega(x)&=&\partial_{\mu}\omega(x)=\frac{\kappa}{\sqrt{6}}\partial_{\mu}\phi(x),\\
	\Omega^{-1}(x)\triangledown_{\mu}\triangledown_{\nu}\Omega(x)&=&\triangledown_{\mu}\triangledown_{\nu}\omega(x)+\partial_{\mu}\omega(x)\partial_{\nu}\omega(x)
	=\frac{\kappa}{\sqrt{6}}\triangledown_{\mu}\triangledown_{\nu}\phi(x)
	+\frac{\kappa^2}{6}\partial_{\mu}\phi(x)\partial_{\nu}\phi(x),~~~~~~~~~~~~~~\\ 
	\Omega^{-1}(x)\Box\Omega(x)&=&\Box\omega(x)+\partial_{\mu}\omega(x)\partial_{\nu}\omega(x)=\frac{\kappa}{\sqrt{6}}\Box\phi(x)
	+\frac{\kappa^2}{6}\partial_{\mu}\phi(x)\partial_{\nu}\phi(x).~~~~~~\eea
	Consequently in terms of ``scalaron'' or ``dilaton'', the Christoffel connections can be recast as:
	\bea \label{eq5aa} \Gamma^{\beta}_{\mu\nu}\Longrightarrow \tilde{\Gamma}^{\beta}_{\mu\nu}&=&\Gamma^{\beta}_{\mu\nu}+
\left[\delta^{\beta}_{\mu}\partial_{\nu}+\delta^{\beta}_{\nu}\partial_{\mu}-g_{\mu\nu}g^{\beta\alpha}\partial_{\alpha}\right]\omega(x)\nonumber\\
&=&\Gamma^{\beta}_{\mu\nu}+\frac{\kappa}{\sqrt{6}}
\left[\delta^{\beta}_{\mu}\partial_{\nu}+\delta^{\beta}_{\nu}\partial_{\mu}-g_{\mu\nu}g^{\beta\alpha}\partial_{\alpha}\right]\phi(x),\\ 
\label{eq5bb} \Gamma^{\nu}_{\mu\nu}\Longrightarrow \tilde{\Gamma}^{\nu}_{\mu\nu}&=&\Gamma^{\nu}_{\mu\nu}+D\partial_{\mu}\omega(x)=\Gamma^{\nu}_{\mu\nu}+\frac{\kappa D}{\sqrt{6}}\partial_{\mu}\phi(x),\\ 
\label{eq5cc} \tilde{\Gamma}^{\beta}_{\mu\nu}\Longrightarrow {\Gamma}^{\beta}_{\mu\nu}&=&\tilde{\Gamma}^{\beta}_{\mu\nu}-
\left[\delta^{\beta}_{\mu}\partial_{\nu}+\delta^{\beta}_{\nu}\partial_{\mu}-\tilde{g}_{\mu\nu}\tilde{g}^{\beta\alpha}\partial_{\alpha}\right]\omega(x)\nonumber\\
&=&\tilde{\Gamma}^{\beta}_{\mu\nu}-\frac{\kappa}{\sqrt{6}}
\left[\delta^{\beta}_{\mu}\partial_{\nu}+\delta^{\beta}_{\nu}\partial_{\mu}-\tilde{g}_{\mu\nu}\tilde{g}^{\beta\alpha}\partial_{\alpha}\right]\phi(x) ,\\ 
\label{eq5dd} \tilde{\Gamma}^{\nu}_{\mu\nu}\Longrightarrow {\Gamma}^{\nu}_{\mu\nu}&=&\tilde{\Gamma}^{\nu}_{\mu\nu}-D\partial_{\mu}\omega(x)=\tilde{\Gamma}^{\nu}_{\mu\nu}-\frac{\kappa D}{\sqrt{6}}\partial_{\mu}\phi(x).\eea
Consequently, the Riemann tensors,  Ricci tensors, and Ricci scalars can be expressed in terms of ``scalaron'' or ``dilaton'' as:
\bea 
	\label{eq6dd} \tilde{R}^{\mu}_{\nu\alpha\beta}\Longrightarrow {R}^{\mu}_{\nu\alpha\beta}&=& \tilde{R}^{\mu}_{\nu\alpha\beta}-
	\left[\delta^{\mu}_{\beta}\tilde{\triangledown}_{\nu}\tilde{\triangledown}_{\alpha}
	-\delta^{\mu}_{\alpha}\tilde{\triangledown}_{\nu}\tilde{\triangledown}_{\beta}+\tilde{g}_{\nu\alpha}
	\tilde{\triangledown}^{\mu}\tilde{\triangledown}_{\beta}-\tilde{g}_{\nu\beta}\tilde{\triangledown}^{\mu}\tilde{\triangledown}_{\alpha}
	\right]\omega(x)\nonumber\\
	&&~~~+\left[\delta^{\mu}_{\alpha}\partial_{\nu}\omega(x)\partial_{\beta}\omega(x)
	-\delta^{\mu}_{\beta}\partial_{\nu}\omega(x)\partial_{\alpha}\omega(x)\nonumber\right.\\ &&\left.~~~~+g_{\nu\beta}\partial^{\mu}\omega(x)
	\partial_{\alpha}\omega(x)
	-g_{\nu\alpha}\partial^{\mu}\omega(x)\partial_{\beta}\omega(x)\right]\nonumber\\
	&&~~~+\left[\delta^{\mu}_{\beta}\tilde{g}_{\nu\alpha}-\delta^{\mu}_{\alpha}\tilde{g}_{\nu\beta}\right]\tilde{g}_{\lambda\eta}
	\partial^{\lambda}\omega(x)\partial^{\eta}\omega(x)\nonumber\\
	&=& \tilde{R}^{\mu}_{\nu\alpha\beta}-
	\frac{\kappa}{\sqrt{6}}\left[\delta^{\mu}_{\beta}\tilde{\triangledown}_{\nu}\tilde{\triangledown}_{\alpha}
	-\delta^{\mu}_{\alpha}\tilde{\triangledown}_{\nu}\tilde{\triangledown}_{\beta}+\tilde{g}_{\nu\alpha}
	\tilde{\triangledown}^{\mu}\tilde{\triangledown}_{\beta}-\tilde{g}_{\nu\beta}\tilde{\triangledown}^{\mu}\tilde{\triangledown}_{\alpha}
	\right]\phi(x)\nonumber\\
	&&~~~+\frac{\kappa^2}{6}\left[\delta^{\mu}_{\alpha}\partial_{\nu}\phi(x)\partial_{\beta}\phi(x)
	-\delta^{\mu}_{\beta}\partial_{\nu}\phi(x)\partial_{\alpha}\phi(x)\nonumber\right.\\ &&\left.~~~~+g_{\nu\beta}\partial^{\mu}\phi(x)
	\partial_{\alpha}\phi(x)
	-g_{\nu\alpha}\partial^{\mu}\phi(x)\partial_{\beta}\phi(x)\right]\nonumber\\
	&&~~~+\frac{\kappa^2}{6}\left[\delta^{\mu}_{\beta}\tilde{g}_{\nu\alpha}-\delta^{\mu}_{\alpha}\tilde{g}_{\nu\beta}\right]\tilde{g}_{\lambda\eta}
	\partial^{\lambda}\phi(x)\partial^{\eta}\phi(x).\\
	\label{eq6ee} \tilde{R}_{\mu\nu}\Longrightarrow {R}_{\mu\nu}&=& \tilde{R}_{\mu\nu}+(D-2)\left[\partial_{\mu}\omega(x)\partial_{\nu}\omega(x)
	-\tilde{g}_{\mu\nu}\partial_{\alpha}\omega(x)\partial^{\alpha}\omega(x)\right]\nonumber\\
	&&~~~+\left[(D-2)\tilde{\triangledown}_{\mu}\tilde{\triangledown}_{\nu}+\tilde{g}_{\mu\nu}\widetilde{\Box}\right]\omega(x)\nonumber\\
	 &=&\tilde{R}_{\mu\nu}+\frac{\kappa^2 (D-2)}{6}\left[\partial_{\mu}\phi(x)\partial_{\nu}\phi(x)
	-\tilde{g}_{\mu\nu}\partial_{\alpha}\phi(x)\partial^{\alpha}\phi(x)\right]\nonumber\\
	&&~~~+\frac{\kappa}{\sqrt{6}}\left[(D-2)\tilde{\triangledown}_{\mu}\tilde{\triangledown}_{\nu}+\tilde{g}_{\mu\nu}\widetilde{\Box}\right]\phi(x),\\
	\label{eq6ff} \tilde{R}\Longrightarrow {R}&=& e^{2\omega(x)}\left[\tilde{R}+2(D-1)
	\widetilde{\Box}\omega(x)\nonumber\right.\\&&\left.~~~~~~~~~~~~~-(D-2)(D-1)\partial_{\alpha}
	\omega(x)\partial_{\beta}\omega(x)\tilde{g}^{\alpha\beta}\right]\nonumber\\
	&=&e^{\frac{2\kappa}{\sqrt{6}}\phi(x)}\left[\tilde{R}+\frac{2\kappa(D-1)}{\sqrt{6}}\widetilde{\Box}\phi(x)
	\nonumber\right.\\&&\left.~~~~~~~~~~~~~-\frac{\kappa^2 (D-2)(D-1)}{6}\partial_{\alpha}\phi(x)\partial_{\beta}\phi(x)\tilde{g}^{\alpha\beta}\right]
	\eea
	Additionaly, the d'Alembertial operator can be expressed in terms of ``scalaron'' or ``dilaton'' as:
	\bea  
	\label{eq7b}\widetilde{\Box}\Longrightarrow{\Box}&=& e^{2\omega(x)}\left[\widetilde{\Box} -(D-2)\tilde{g}^{\mu\nu}\partial_{\mu}\omega(x)\partial_{\nu}\right]=
	e^{\frac{2\kappa}{\sqrt{6}}\phi(x)}\left[\widetilde{\Box} -\frac{\kappa(D-2)}{\sqrt{6}}\tilde{g}^{\mu\nu}
	\partial_{\mu}\phi(x)\partial_{\nu}\right].~~~~~~~~~\eea
	Finally, the Einstein tensor is transformed as:
	\bea\label{eq7v4} \tilde{G}_{\mu\nu}\Longrightarrow {G}_{\mu\nu}&=&\tilde{G}_{\mu\nu}+\left(\frac{D-2}{2}\right)\left[\partial_{\mu}\omega(x)\partial_{\nu}\omega(x)+\left(\frac{D-3}{2}\right)
	\partial_{\alpha}\omega(x)\partial^{\alpha}\omega(x)\tilde{g}_{\mu\nu}\right]\nonumber\\
	&& ~~~~~~~~~~~~~~~~+(D-2)\left[\tilde{\triangledown}_{\mu}\tilde{\triangledown}_{\nu}-\tilde{g}_{\mu\nu}\widetilde{\Box}\right]\omega(x)\nonumber \\
	&=&\tilde{G}_{\mu\nu}+\frac{\kappa^2}{6}\left(\frac{D-2}{2}\right)\left[\partial_{\mu}\phi(x)\partial_{\nu}\phi(x)+\left(\frac{D-3}{2}\right)
	\partial_{\alpha}\phi(x)\partial^{\alpha}\phi(x)\tilde{g}_{\mu\nu}\right]\nonumber\\
	&& ~~~~~~~~~~~~~~~~+\frac{\kappa}{\sqrt{6}}(D-2)\left[\tilde{\triangledown}_{\mu}\tilde{\triangledown}_{\nu}-\tilde{g}_{\mu\nu}\widetilde{\Box}\right]\phi(x).\eea
	We use the results for $D=4$ to study the consequences in the context of dark matter.
	\subsection{Conformal transformations in matter sector}
	\label{s2b}
	Let us assume that  matter is minimally coupled with the gravity sector. In such a case, in an arbitrary $D$ dimensional space-time, the action can be written as:
	\be S_{M}=\int d^{D}x\sqrt{-g}{\cal L}_{M},\ee
	which is invariant under the conformal transformation in the metric, as mentioned earlier. In our present context, in $D=4$, we have taken the matter sector to be SM i.e. ${\cal L}_{M}={\cal L}_{SM}$.
	Under this conformal transformation, the energy-momentum stress tensor transforms as:
	\bea \label{eq9a} \tilde{T}^{\mu\nu}_{M}&=&\frac{2}{\sqrt{-\tilde{g}}}\frac{\delta}{\delta\tilde{g}_{\mu\nu}}\left(\sqrt{-\tilde{g}}\tilde{\cal L}_{M}\right)=\Omega^{-D-2}(x){T}^{\mu\nu}_{M}
	=e^{-(D+2)w(x)}{T}^{\mu\nu}_{M}=e^{-\frac{\kappa}{\sqrt{6}}
	(D+2)\phi(x)}{T}^{\mu\nu}_{M},~~~~~~~~~~~~\\
	\label{eq9b} \tilde{T}^{\mu}_{\nu,M}&=& \tilde{T}^{\mu\alpha}_{M}\tilde{g}_{\alpha\nu}=\Omega^{-D}(x){T}^{\mu}_{\nu,M}=e^{-Dw(x)}{T}^{\mu}_{\nu,M}=e^{-\frac{\kappa}{\sqrt{6}}D\phi(x)}{T}^{\mu}_{\nu,M},\\ 
	\label{eq9c} \tilde{T}_{\mu\nu,M}&=& \tilde{T}^{\alpha\beta}_{M}\tilde{g}_{\alpha\mu}\tilde{g}_{\beta\nu}=\Omega^{-D+2}(x){T}_{\mu\nu,M}=e^{(2-D)w(x)}{T}_{\mu\nu,M}=e^{\frac{\kappa}{\sqrt{6}}(2-D)\phi(x)}{T}_{\mu\nu,M},\\ 
	\label{eq9d} \tilde{T}_{M}&=&T^{\mu}_{\mu,M}=\Omega^{-D}(x){T}^{\mu}_{\mu,M}=e^{-Dw(x)}{T}^{\mu}_{\mu,M}=e^{-\frac{\kappa}{\sqrt{6}}D\phi(x)}{T}^{\mu}_{\mu,M},\eea
	where $\tilde{\cal L}_{M}$ is the energy-momentum stress tensor in {\it Einstein frame} and this is related to the {\it Jordan frame} via the following transformation rule:
	\bea \label{eq10} \tilde{\cal L}_{M}=\Omega^{-D}(x){\cal L}_{M}=e^{-Dw(x)}{\cal L}_{M}=e^{-\frac{\kappa}{\sqrt{6}}D\phi(x)}{\cal L}_{M}.\eea
	Using the the fact that the matter sector is governed by a perfect fluid and the structural form of the conformal transformation in the metric,
	one can show that the density and pressure can be transformed in the {\it Einstein frame} as:
	\bea \label{eq11a} \tilde{\rho}&=&\Omega^{-D}(x)\rho=e^{-Dw(x)}\rho=e^{-\frac{\kappa}{\sqrt{6}}D\phi(x)}\rho,\\
	\label{eq11b} \tilde{p}&=&\Omega^{-D}(x)p=e^{-Dw(x)}p=e^{-\frac{\kappa}{\sqrt{6}}D\phi(x)}p,\eea
	where $(\rho,p)$ and $(\tilde{\rho},\tilde{p})$ are the density and pressure of the matter content in {\it Jordan} and {\it Einstein} frame respectively. The results clearly show that if we impose conservation of the energy-momentum stress tensor 
        in one frame then in the other conformally connected frame it is no longer conserved. Only if we assume that in both the frames matter content is governed by the traceless tensor, then  
	conservation holds good in both the frames simultaneously. But for a general matter content this may not always be the case. For example, in the $D=4$ version of the Effective Field Theory discussed in this paper, we
	assume that the matter content is governed by the well known SM fields in the {\it Jordan frame}. But after applying the conformal transformation in the metric, the conformal coupling factor becomes
	\be \Omega^{-4}(x)=e^{-4\omega(x)}=e^{-\frac{4\kappa}{\sqrt{6}}\phi(x)},\ee
	or more precisely, the ``scalaron''or the ``dilaton''
	field is interacting with the SM matter fields in the {\it Einstein frame}, which will act as the primary source of generating a scalar dark matter candidate
	from an extended theory of gravity.

%%
%%%%%%%%%%%%%%%%%%%%%%%
\section{Appendix B:~Thermally averaged annihilation cross-section}
\label{s8}
Here we outline the annihilation cross section for the processes contributing to the relic density.
\begin{figure}[!htb]
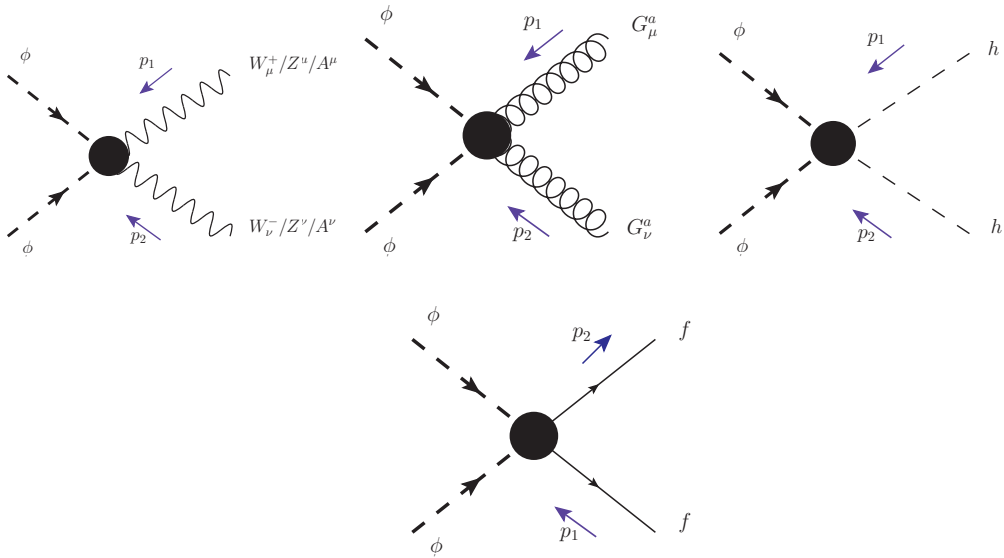

	   \includegraphics[width=0.3\textwidth]{boson_boson.pdf}
           \includegraphics[width=0.3\textwidth]{glue_glue.pdf}
           \includegraphics[width=0.3\textwidth]{higgs_higgs2.pdf}
           \begin{center}
           \includegraphics[width=0.3\textwidth]{fermion_fermion.pdf}
           \end{center}
           \caption{Effective processes contributing to relic density of dark matter.}
           \label{fig10}
	  \end{figure}  
           \bea
            \langle\sigma v\rangle_{NR_{G_\mu G^\mu}}&=&\frac{32 M^2}{\pi \Lambda^4_{UV}}+ \frac{16 M^2}{\pi \Lambda^4_{UV}}v^2= a_{NR_{G_\mu G^\mu}}(\Lambda_{UV}, M)+ b_{NR_{G_\mu G^\mu}}(\Lambda_{UV}, M)v^2~~~\\
            \langle\sigma v\rangle_{NR_{W_\mu W^\mu}}&=&\frac{\sqrt{1-\frac{M_W^2}{M^2}}}{32 \pi M^2} \left(\frac{64 M^4}{\Lambda^4_{UV}}+\frac{64 M^8}{\Lambda^4_{UV} M_W^4}-\frac{128 M^6}{\Lambda^4_{UV}M_W^2}
                                                       +\frac{8 M_W^4}{\Lambda^4_{UV}}\right)\nonumber\\
                                                       & &~~~~+\frac{\sqrt{1-\frac{M_W^2}{M^2}}}{32 \pi M^2}
                                                       \left(\frac{32 M^4}{\Lambda^4_{UV}}+\frac{64 M^8}{\Lambda^4_{UV}M_W^4 }-\frac{96 M^6}{\Lambda^4_{UV} M_W^2}\right)v^2 \nonumber\\
            &=& a_{NR_{W_\mu W^\mu}}(\Lambda_{UV}, M)+ b_{NR_{W_\mu W^\mu}}(\Lambda_{UV}, M)v^2\\
              \langle\sigma v\rangle_{NR_{Z_\mu Z^\mu}}&=&\frac{\sqrt{1-\frac{M_Z^2}{M^2}}}{32 \pi M^2} \sin^4\theta_W\left(\frac{64 M^4}{\Lambda^4_{UV}}+\frac{64 M^8}{\Lambda^4_{UV} M_Z^4}-\frac{128 M^6}{\Lambda^4_{UV}M_Z^2}
                                                       +\frac{8 M_Z^4}{\Lambda^4_{UV}}\right)\nonumber\\
                                                       & &~~~~+\frac{\sqrt{1-\frac{M_Z^2}{M^2}}}{32 \pi M^2}
                                                       \sin^4\theta_W\left(\frac{32 M^4}{\Lambda^4_{UV}}+\frac{64 M^8}{\Lambda^4_{UV}M_Z^4 }-\frac{96 M^6}{\Lambda^4_{UV} M_Z^2}\right)v^2\nonumber\\
            &=& a_{NR_{Z_\mu Z^\mu}}(\Lambda_{UV}, M)+ b_{NR_{Z_\mu Z^\mu}}(\Lambda_{UV}, M)v^2\\
            \langle\sigma v\rangle_{NR_{A_\mu A^\mu}}&=&\frac{4 M^2 \cos^4(\theta_W)}{\pi \Lambda^4_{UV}}+ \frac{2 M^2 \cos^4(\theta_W)}{\pi \Lambda^4_{UV}}v^2\nonumber\\
            &=& a_{NR_{A_\mu A^\mu}}(\Lambda_{UV}, M)+ b_{NR_{A_\mu A^\mu}}(\Lambda_{UV}, M)v^2\\
           \langle\sigma v\rangle_{NR_{hh}} &=& \frac{\sqrt{1-\frac{M_h^2}{M^2}}}{32 \pi M^2} \left(\frac{64 M^4}{\Lambda^4_{UV}}+\frac{32 M^4}{\Lambda^4_{UV}}v^2\right)=
           a_{NR_{hh}}(\Lambda_{UV}, M)+ b_{NR_{hh}}(\Lambda_{UV}, M)v^2~~~~~~~\\
           \langle\sigma v\rangle_{NR_{ff}}&=& \frac{\sqrt{1-\frac{M_f^2}{M^2}}}{32 \pi M^2} \left(\frac{4 M^2 M_f^2}{\Lambda^4_{UV}}-\frac{4 M_f^4}{\Lambda^4_{UV}}+\frac{ M^2 M_f^2}{\Lambda^4_{UV}}v^2\right)\nonumber\\
            &=& a_{NR_{ff}}(\Lambda_{UV}, M)+ b_{NR_{ff}}(\Lambda_{UV}, M)v^2.
          \eea where $f$ can be any fermion channel which is kinematically allowed.
          Here the expression for $a$ and $b$ for the individual processes are given by:
          \bea
            a_{NR_{G_\mu G^\mu}}(\Lambda_{UV}, M)&=&\frac{32 M^2}{\pi \Lambda^4_{UV}},\\
            a_{NR_{W_\mu W^\mu}}(\Lambda_{UV}, M)&=&\frac{\sqrt{1-\frac{M_W^2}{M^2}}}{32 \pi M^2} \left(\frac{64 M^4}{\Lambda^4_{UV}}+\frac{64 M^8}{\Lambda^4_{UV} M_W^4}-\frac{128 M^6}{\Lambda^4_{UV}M_W^2}
                                                       +\frac{8 M_W^4}{\Lambda^4_{UV}}\right),\\
              a_{NR_{Z_\mu Z^\mu}}(\Lambda_{UV}, M)&=&\frac{\sqrt{1-\frac{M_Z^2}{M^2}}}{32 \pi M^2} \sin^4\theta_W\left(\frac{64 M^4}{\Lambda^4_{UV}}+\frac{64 M^8}{\Lambda^4_{UV} M_Z^4}-\frac{128 M^6}{\Lambda^4_{UV}M_Z^2}
                                                       +\frac{8 M_Z^4}{\Lambda^4_{UV}}\right),~~~~~~~\\
            a_{NR_{A_\mu A^\mu}}(\Lambda_{UV}, M)&=&\frac{4 M^2 \cos^4(\theta_W)}{\pi \Lambda^4_{UV}},\\
           a_{NR_{hh}}(\Lambda_{UV}, M) &=& \frac{\sqrt{1-\frac{M_h^2}{M^2}}}{32 \pi M^2} \frac{64 M^4}{\Lambda^4_{UV}},\\
           a_{NR_{ff}}(\Lambda_{UV}, M)&=& \frac{\sqrt{1-\frac{M_f^2}{M^2}}}{32 \pi M^2} \left(\frac{4 M^2 M_f^2}{\Lambda^4_{UV}}-\frac{4 M_f^4}{\Lambda^4_{UV}}\right),
          \eea
          \bea
            b_{NR_{G_\mu G^\mu}}(\Lambda_{UV}, M)&=&\frac{16 M^2}{\pi \Lambda^4_{UV}}~~~\\
            b_{NR_{W_\mu W^\mu}}(\Lambda_{UV}, M)&=&\frac{\sqrt{1-\frac{M_W^2}{M^2}}}{32 \pi M^2}
                                                       \left(\frac{32 M^4}{\Lambda^4_{UV}}+\frac{64 M^8}{\Lambda^4_{UV}M_W^4 }-\frac{96 M^6}{\Lambda^4_{UV} M_W^2}\right)\\
             b_{NR_{Z_\mu Z^\mu}}(\Lambda_{UV}, M)&=&\frac{\sqrt{1-\frac{M_Z^2}{M^2}}}{32 \pi M^2}
                                                       \sin^4\theta_W\left(\frac{32 M^4}{\Lambda^4_{UV}}+\frac{64 M^8}{\Lambda^4_{UV}M_Z^4 }-\frac{96 M^6}{\Lambda^4_{UV} M_Z^2}\right)\\
           b_{NR_{A_\mu A^\mu}}(\Lambda_{UV}, M)&=&\frac{2 M^2 \cos^4(\theta_W)}{\pi \Lambda^4_{UV}}\\
           b_{NR_{hh}}(\Lambda_{UV}, M) &=& \frac{\sqrt{1-\frac{M_h^2}{M^2}}}{32 \pi M^2} \frac{32 M^4}{\Lambda^4_{UV}}\\
           b_{NR_{ff}}(\Lambda_{UV}, M)&=& \frac{\sqrt{1-\frac{M_f^2}{M^2}}}{32 \pi M^2} \left(\frac{4 M^2 M_f^2}{\Lambda^4_{UV}}-\frac{4 M_f^4}{\Lambda^4_{UV}}+\frac{ M^2 M_f^2}{\Lambda^4_{UV}}v^2\right).
          \eea
           Therefore, summing up all the contributions, we get \bea \langle\sigma v\rangle_{NR}&=&  \langle\sigma v\rangle_{NR_{G_\mu G^\mu}}+\langle\sigma v\rangle_{NR_{W_\mu W^\mu}}+\langle\sigma v\rangle_{NR_{Z_\mu Z^\mu}}
           +\langle\sigma v\rangle_{NR_{A_\mu A^\mu}}+\langle\sigma v\rangle_{NR_{hh}}+\langle\sigma v\rangle_{NR_{ff}}\nonumber\\ 
           &=&a(\Lambda_{UV}, M)+ b(\Lambda_{UV}, M)v^2,\eea where $a(\Lambda_{UV}, M)$ and $b(\Lambda_{UV}, M)$ is defined as:
           \begin{eqnarray}
            a(\Lambda_{UV}, M)&=&a_{NR_{G_\mu G^\mu}}(\Lambda_{UV}, M)+a_{NR_{W_\mu W^\mu}}(\Lambda_{UV}, M)+a_{NR_{Z_\mu Z^\mu}}(\Lambda_{UV}, M)
           \nonumber\\&&~~~~~~~~~~~~~+a_{NR_{A_\mu A^\mu}}(\Lambda_{UV}, M)+a_{NR_{hh}}(\Lambda_{UV}, M)+a_{NR_{ff}}(\Lambda_{UV}, M)\nonumber\\ &=&\frac{32 M^2}{\pi \Lambda^4_{UV}}+\frac{\sqrt{1-\frac{M_W^2}{M^2}}}{32 \pi M^2}\left(\frac{64 M^4}{\Lambda^4_{UV}}+\frac{64 M^8}{\Lambda^4_{UV} M_W^4}-\frac{128 M^6}{\Lambda^4_{UV}M_W^2} +\frac{8 M_W^4}{\Lambda^4_{UV}}\right)+\nonumber \\
                              & &\frac{\sqrt{1-\frac{M_Z^2}{M^2}}}{32 \pi M^2} \sin^4\theta_W\left(\frac{64 M^4}{\Lambda^4_{UV}}+\frac{64 M^8}{\Lambda^4_{UV} M_Z^4}-\frac{128 M^6}{\Lambda^4_{UV}M_Z^2} +\frac{8 M_Z^4}{\Lambda^4_{UV}}\right)+\nonumber\\
                              & &\frac{4 M^2 \cos^4(\theta_W)}{\pi \Lambda^4_{UV}}+ \frac{\sqrt{1-\frac{M_h^2}{M^2}}}{32 \pi M^2} \left(\frac{64 M^4}{\Lambda^4_{UV}}\right)+\frac{\sqrt{1-\frac{M_f^2}{M^2}}}{32 \pi M^2} 
                              \left(\frac{4 M^2 M_f^2}{\Lambda^4_{UV}}-\frac{4 M_f^4}{\Lambda^4_{UV}}\right),~~~~~~~~~~~\\
            b(\Lambda_{UV}, M)&=&b_{NR_{G_\mu G^\mu}}(\Lambda_{UV}, M)+b_{NR_{W_\mu W^\mu}}(\Lambda_{UV}, M)+b_{NR_{Z_\mu Z^\mu}}(\Lambda_{UV}, M)
           \nonumber\\&&~~~~~~~~~~~~~+b_{NR_{A_\mu A^\mu}}(\Lambda_{UV}, M)+b_{NR_{hh}}(\Lambda_{UV}, M)+b_{NR_{ff}}(\Lambda_{UV}, M)\nonumber\\&=& \frac{16 M^2}{\pi \Lambda^4_{UV}}+\frac{\sqrt{1-\frac{M_W^2}{M^2}}}{32 \pi M^2} \left(\frac{32 M^4}{\Lambda^4_{UV}}+\frac{64 M^8}{\Lambda^4_{UV}M_W^4 }-\frac{96 M^6}{\Lambda^4_{UV} M_W^2}\right)+\nonumber \\
                              & &\frac{\sqrt{1-\frac{M_Z^2}{M^2}}}{32 \pi M^2}\sin^4\theta_W\left(\frac{32 M^4}{\Lambda^4_{UV}}+\frac{64 M^8}{\Lambda^4_{UV}M_Z^4 }-\frac{96 M^6}{\Lambda^4_{UV} M_Z^2}\right)\nonumber \\
                              & &\frac{2 M^2 \cos^4(\theta_W)}{\pi \Lambda^4_{UV}}+ \frac{\sqrt{1-\frac{M_h^2}{M^2}}}{32 \pi M^2} \left(\frac{32 M^4}{\Lambda^4_{UV}}\right)+
                              \frac{\sqrt{1-\frac{M_f^2}{M^2}}}{32 \pi M^2} \left(\frac{ M^2 M_f^2}{\Lambda^4_{UV}}\right).
           \end{eqnarray}
           
%%%%%%%%%%%%%%%%%
%%%%%%%%%%%%%%%%%%%%%%%
\section{Appendix C:~Effective potential construction for dark matter self interaction}
\label{s9}
In this section we discuss about the effective potential construction necessarily required for dark matter self interaction. Using the results of this section derived from modified gravity
-(1)$f(R)$ gravity, (2) non-minimally coupled gravity theory we further constrain the parameters of the modified gravity theories.
\subsection{Case I: For $f(R)$ gravity}

\subsubsection{A. For $n=2$}
In this case $f(R)$ is given by:
\bea f(R)=a R+b R^2,\eea
where we set $a=1$ to have consistency with the Einstein gravity at the leading order and in this case $b$ is the only parameter that has to be constrained from dark matter self interaction for $n=2$ case.
Additionally it is important to note that the mass dimension of $b$ for $n=2$ case is $-2$.

In the present context, the effective potential can be expressed as:
	\be\begin{array}{lll}\label{eq17}
		\displaystyle   V(\phi)=
			\frac{\Lambda^2_{UV}}{4b}e^{-\frac{4}{\sqrt{6}}\frac{\phi(x)}{\Lambda_{UV}}}\left(e^{\frac{2}{\sqrt{6}}\frac{\phi(x)}{\Lambda_{UV}}}-a\right)^2\,.
	\end{array}\ee
To further study the constraint on the model parameters, one can expand the effective potential by respecting the ${\cal Z}_{2}$ symmetry as:
\bea \label{fg12} V(\phi)=V_0 + \frac{V^{''}_{0}}{2!}\phi^2 + \frac{V^{''''}_{0}}{4!}\phi^4 + \cdots,\eea
where the Taylor expansion coefficients are given by:
\bea V_{0}&=&0,\\
V^{''}_{0}&=&\frac{1}{3b},\\
V^{''''}_{0}&=&24 \lambda=\frac{14}{9b\Lambda^2_{UV}}.\eea

\subsubsection{B. For $n>2$}
In this case $f(R)$ is fiven by:
\bea f(R)=a R+b R^n,\eea
where for physical consistency, we set $a\neq 1$ and in this case, $a$ and $b$ are the parameters to be constrained from dark matter self interaction for $n>2$ case.
Here it is important to note that, for the further numerical estimation we set $n=3$. Additionally it is important to note that the mass dimension of $b$ for $n=3$ case is $-4$.

In the present context the effective potential can be expressed as:
	\be\begin{array}{lll}\label{eq17}
		\displaystyle   V(\phi)=
			A\Lambda^2_{UV}~e^{-\frac{4}{\sqrt{6}}\frac{\phi(x)}{\Lambda_{UV}}}
			\left(e^{\frac{2}{\sqrt{6}}\frac{\phi(x)}{\Lambda_{UV}}}-a\right)^{B}\,.
	\end{array}\ee
	where $A$ and $B$ are defined as:
	\bea A&=&\frac{b(n-1)}{(nb)^{\frac{n}{n-1}}},\\
	B&=&\frac{n}{n-1}.\eea
To further study the constraint on the model parameters, one can expand the effective potential by respecting the ${\cal Z}_{2}$ symmetry as:
\bea \label{fg13} V(\phi)=V_0 + \frac{V^{''}_{0}}{2!}\phi^2 + \frac{V^{''''}_{0}}{4!}\phi^4 + \cdots,\eea
where the Taylor expansion coefficients are given by:
\bea V_{0}&=&0,\\
V^{''}_{0}&=&\frac{2A}{3}(1-a)^{B-2}\left[4a^2+(B-2)^2+a(3B-8)\right],\\
V^{''''}_{0}&=&24 \lambda=\frac{4A}{9\Lambda^2_{UV}}(1-a)^{B-4}\left[16a^4+(B-2)^4+a^2(B-4)(7B-24)\right.\nonumber\\ &&\left.~~~~~~~~~~~~~~~~~~~~~~+a^3(15B-64)+a(-64+B(63+2B(B-10)))\right].~~~~~~~~\eea
Therefore,
\begin{equation}
 \lambda= \frac{0.0004+a\left[-0.0552+a\left(0.2405+(0.1140a-0.2958)a\right)\right]}{(1-a)^{2.5} b^{0.5} \Lambda^2_{UV}} \nonumber
\end{equation}
\subsection{Case II: For non-minimally couples gravity with $\xi\neq 1/6,\frac{\phi}{\Lambda_{UV}}>>\frac{1}{\xi}$}
Here we will discuss the situation where $\xi\neq 1/6,\frac{\phi}{\Lambda_{UV}}>>\frac{1}{\xi}$ and the effect of the non-minimal coupling $\xi$ can be visualized prominantly as it 
 couples to the SM sector. The other case $\xi\neq 1/6,\frac{\phi}{\Lambda_{UV}}<<\frac{1}{\xi}$ is not relevant in the present context as in this case the effect of the non-minimal coupling $\xi$ can be 
 neglected and SM sector couples to gravity minimally. In $\xi\neq 1/6,\frac{\phi}{\Lambda_{UV}}>>\frac{1}{\xi}$ case the only parameter for the modified gravity theory is the non-minimal coupling $\xi$ for the 
 given value of dimensionless coefficients $C_{2}(g)$ and $C_{4}(g)$ and here we 
 will constrain $\xi$ using the constraint from dark matter self interaction. For the sake of simplicity we set $C_{2}(g)\sim C_{4}(g)\sim {\cal O}(1)$.
 
In the present context the effective potential can be expressed as:
	\be\begin{array}{lll}\label{eq17}
		\displaystyle   V(\phi)=
			e^{-\frac{4{\phi}}{\sqrt{6}\Lambda_{UV}}}\sum^{\infty}_{\Delta_{\alpha}=0}C_{\Delta_{\alpha}}
		(g)\frac{\Lambda^{4}_{UV}}{\xi^{\frac{\Delta_{\alpha}}{2}}}\left(e^{\frac{2{\phi}}
		{\sqrt{6}\Lambda_{UV}}}-1\right)^{\frac{\Delta_{\alpha}}{2}}\,.
	\end{array}\ee
	Here for numerical study we trucate the above series at $\Delta_{\alpha}=4$ and applying
	${\cal Z}_{2}$ symmetry of the effective potential one can write down the expression:
	\bea\label{eq17vb}
		\displaystyle   V(\phi)&=&
			e^{-\frac{4{\phi}}{\sqrt{6}\Lambda_{UV}}}\sum_{\Delta_{\alpha}=0,2,4}C_{\Delta_{\alpha}}
		(g)\frac{\Lambda^{4}_{UV}}{\xi^{\frac{\Delta_{\alpha}}{2}}}\left(e^{\frac{2{\phi}}
		{\sqrt{6}\Lambda_{UV}}}-1\right)^{\frac{\Delta_{\alpha}}{2}}\,\nonumber\\
		&=&\left[A+B~e^{-\frac{2{\phi}}{\sqrt{6}\Lambda_{UV}}}+C~e^{-\frac{4{\phi}}
		{\sqrt{6}\Lambda_{UV}}}\right],\eea
	where $A$, $B$ and $C$ is given by:
	\bea A&=& \Lambda^4_{UV} \frac{C_{4}(g)}{\xi},\\ 
	B&=&\Lambda^4_{UV} \frac{\left(C_{2}(g)-\frac{2C_{4}(g)}{\xi}\right)}{\xi},\\
	C&=&\Lambda^4_{UV} \left(C_{0}(g)-\frac{C_{2}(g)}{\xi}+\frac{C_{4}(g)}{\xi^2}\right).\eea

To further study the constraint on the model parameters, one can expand the effective potential by respecting the ${\cal Z}_{2}$ symmetry as:
\bea \label{fg13} V(\phi)=V_0 + \frac{V^{''}_{0}}{2!}\phi^2 + \frac{V^{''''}_{0}}{4!}\phi^4 + \cdots,\eea
where the Taylor expansion coefficients are given by:
\bea V_{0}&=&0,\\
V^{''}_{0}&=&\frac{B+4C}{9\Lambda^2_{UV}},\\
V^{''''}_{0}&=&24 \lambda=\frac{B+16C}{81\Lambda^4_{UV}}.\eea
For $C_{0}(g)\sim C_{2}(g)\sim C_{4}(g)\sim{\cal O}(1)$, we get the following expression for the self interaction parameter 
\begin{equation}
 \lambda=\frac{14+\xi(16\xi-15)}{1944 \xi^2}.
\end{equation}

%%%%%%%%%%%%%%%%%%%%%%%%%%%%%%%%%%%%%%%%%%%%%%%%%%%%%%%%%%%%%%%%%%%%%%%%%%%%%%%	

\end{document}